\documentclass[12pt]{article}
\newcommand{\be}{\begin{equation}}
\newcommand{\ee}{\end{equation}}

\newcommand{\bi}{\begin{itemize}}
\newcommand{\ei}{\end{itemize}}
\newcommand{\bea}{\begin{eqnarray}}
\newcommand{\eea}{\end{eqnarray}}
\newcommand{\ba}{\begin{array}}
\newcommand{\ea}{\end{array}}

\usepackage[left=2.50cm, right=2.50cm, top=2.50cm, bottom=2.50cm]{geometry}
\usepackage[utf8]{inputenc}
\usepackage{cancel}
\usepackage[english]{babel}
\usepackage{physics}
\usepackage{hyperref}
\usepackage{amsthm}
\usepackage{mathrsfs}
\usepackage{mathtools}
\usepackage{graphicx}
\usepackage{changepage}
\usepackage{amssymb,amsmath}
\usepackage{verbatim}
\usepackage{empheq}
\usepackage{xcolor}
\usepackage{tikz}
\usepackage{float}
\usepackage{multirow}
\usepackage{multicol}
\usepackage{amsmath}
\usepackage{centernot}
\usepackage[hang,flushmargin]{footmisc} 
\usepackage[normalem]{ulem}

\usepackage[backend=bibtex,style=ieee,dashed=false,citestyle=numeric-comp,eprint=false, url=false, doi=true,isbn=false]{biblatex}

\DeclareFieldFormat[article]{title}{}
\bibliography{references2}

\usetikzlibrary{tikzmark}
\numberwithin{equation}{section}
\hypersetup{hidelinks}

\begin{document}
\par
\bigskip
\Large
\noindent
{\bf 
Fractonic self-duality and covariant magnetic fractons
\\
\rm
\normalsize

\hrule

\vspace{1cm}

\large
\noindent
{\bf Erica Bertolini$^{1,a}$}, {\bf Giandomenico Palumbo$^{1,b}$}

\par

\small

\noindent$^1$ School of Theoretical Physics, Dublin Institute for Advanced Studies, 10 Burlington Road, Dublin 4, D04 C932, Ireland.

\smallskip

\smallskip

\vspace{1cm}

\noindent{\tt Abstract:} \\
Fractons, excitations with restricted mobility, have emerged as a novel paradigm in high-energy and condensed matter physics, revealing deep connections to gauge theories and gravity. Here, we propose a tensorial generalization of electromagnetic duality using a doubled-potential framework with two symmetric tensor gauge fields. This approach symmetrically describes electric and magnetic sectors, introducing covariant magnetic fractons with reduced mobility and supporting a genuine fractonic self-duality. Notably, we find the absence of a Witten-like effect in the covariant fractonic case, precluding the emergence of fractonic dyons.

\vspace{\fill}

\noindent{\tt Keywords:} \\
Quantum field theory, symmetric tensor gauge field theory, fractons, dualities.

\vspace{1cm}

\hrule
\noindent{\tt E-mail:
$^a$ebertolini@stp.dias.ie,
$^b$giandomenico.palumbo@gmail.com.
}
\newpage

\section{Introduction}

Dualities play a central role in the non-perturbative aspects of both free and interacting quantum field theories \cite{Polchinski:2014mva}. Notable examples in three and four dimensions, such as fermion-fermion \cite{Son:2015xqa,Wang:2015qmt,Metlitski:2015eka,Palumbo:2019tmg,Grushin:2020kdo}, fermion-boson \cite{Polyakov:1988md,Fradkin:1994tt,Karch:2016sxi,Palumbo:2014lqa,Palumbo:2013rb}, boson-vortex \cite{Dasgupta:1981zz,Fisher:1989dnp,Franz:2006gb}, S- \cite{Witten:1995gf,Dimofte:2011jd,Metlitski:2015yqa}, and T-dualities \cite{Alvarez:1994dn,Bergshoeff:2018yvt,Mathai:2015raa}, have revealed intricate relationships between seemingly distinct quantum field theories. Recently, fractons have emerged as a novel paradigm in both high-energy and condensed matter physics. Originally introduced in non-relativistic systems \cite{Pretko:2016kxt,Pretko:2017fbf,Pretko:2019omh,Gromov:2017vir,Glodkowski:2022xje,Grosvenor:2021rrt,Perez:2022kax,Figueroa-OFarrill:2023vbj,Seiberg:2020wsg,Gorantla:2022ssr} inspired by specific microscopic condensed matter models \cite{Chamon:2004lew,Haah:2011drr,Vijay:2016phm,Prem:2017kxc,You:2019bvu}, fractons have since been extended into a fully covariant and relativistic framework \cite{Blasi:2022mbl,Bertolini:2022ijb,Bertolini:2023juh,Bertolini:2023sqa,Bertolini:2024yur,Bertolini:2024apg,Bertolini:2025qcy}, also highlighting intriguing connections to gravity, curved spacetimes \cite{Slagle:2018kqf,Bidussi:2021nmp,Pena-Benitez:2021ipo,Pena-Benitez:2023aat,Jain:2021ibh, Afxonidis:2023pdq,Tsaloukidis:2023bvz,Hartong:2024hvs} and electromagnetism. Indeed fracton models share with gravity the tensorial nature of its gauge field, which is that of a rank-2 symmetric tensor transforming under a peculiar case of infinitesimal diffeomorphisms known as longitudinal diffeomorphisms \cite{Dalmazi:2020xou}
	\be
	\delta A_{\mu\nu}=\partial_\mu\partial_\nu\lambda\ .
	\ee
The connection with electromagnetism emerges for instance from the equations of motion (EoM), which reflect a form of higher-rank Maxwell-like theory \cite{Pretko:2016lgv,Pretko:2016kxt,Bertolini:2022ijb}, which encode the main feature of fractons, $i.e.$ immobility, in the Gauss-like constraint. The restricted motion of fracton models is typically associated to dipole/multipole conservations \cite{Pretko:2016lgv,Pretko:2016kxt,Gromov:2018nbv,Caddeo:2022ibe}, which is thus a fingerprint of a fractonic behaviour. The prototypical fracton gauge theory, the scalar charge theory \cite{Pretko:2016lgv,Pretko:2016kxt}, is characterised by a symmetric electric tensor field $E^{ij}(x)$ being the conjugate momentum of the gauge field $A_{ij}(x)$, whose Gauss constraint is given by
	\be
	\partial_i\partial_jE^{ij}=\rho\ ,
	\ee
where $\rho(x)$ is fractonic matter. This relation implies the conservation of the dipole moment $x^i\rho(x)$ as
	\be
	\int dV x^i\rho=\int dV x^i\partial_j\partial_kE^{jk}=-\int dV\partial_jE^{ij}=0,
	\ee
when integrated over a volume (up to boundary terms), with the consequence that single charges, in isolation, cannot move \cite{Pretko:2016lgv,Pretko:2016kxt}. To different Gauss constraints correspond different kinds of multipole conservations and mobilities \cite{Bertolini:2024jen,Iaconis:2020zhc,Osborne:2021mej, Gromov:2018nbv}. These excitations, characterized by immobility or restricted mobility, have broadened our understanding of gauge theories with tensorial fields and have significant implications for many-body physics with constrained dynamics. This higher rank electromagnetic (EM) analogy thus allows to take insights from standard electromagnetism, in order uncover new features for fracton models. This is the case for instance regarding magnetic charges: the ``scalar charge theory of fractons'' \cite{Pretko:2016lgv,Pretko:2016kxt}, for example, has magnetic fractonic charges that are introduced by hand. In the case of electromagnetism a similar thing happens, unless one consider a ``doubled-potential'' approach \cite{Cabibbo:1962td,Salam:1966bd,Zwanziger:1970hk,Blagojevic:1985sh,Singleton:1996hgp,CasteloFerreira:2005ej}. This doubled-potential approach was originally introduced in \cite{Cabibbo:1962td,Salam:1966bd,Zwanziger:1970hk} and is based on the idea of imposing the EM duality at the level of the action and not only at the level of the EoM. Under the EM duality the following transformations of the electric and magnetic fields
	\be\label{EBdual2}
	E^a\to B^a\quad;\quad B^a\to-E^a\ ,
	\ee
leave invariant the Maxwell equations in vacuum, but not its action, in the standard case. This invariance is also broken at the level of the equations of motion in the presence of either a curved background \cite{Agullo:2014yqa,Palumbo:2016nku} or magnetic currents and magnetic monopoles. In other words, in flat spacetime, Maxwell equations with magnetic monopoles cannot be derived from the usual action unless a second gauge field is introduced which can be then associated to a magnetic sector \cite{Bunster:2011qp,Bunster:2012km}. It is therefore interesting to investigate whether such a duality can be extended to fractonic models as well, also allowing for the existence of magnetic fractonic charges. Beyond quantum field theory, magnetic monopoles have been observed in certain quantum spin liquids, such as spin ice \cite{Castelnovo:2007qi}. Covariant magnetic fractons could, therefore, provide insights into a covariant generalization of quantum spin liquids that host fractonic magnetic excitations \cite{Yan:2019qch,Zhang:2021ehb}. With this aim in mind, in this work, we propose a simple yet non-trivial generalization of the covariant fractonic model \cite{Bertolini:2022ijb}, which supports an electromagnetic-like duality at the level of the action, and the existence of magnetic fractonic excitations. \\

The paper is organized as follows. In Section \ref{EMduality} we analyze the standard EM duality within the doubled-potential formalism. In particular in Section \ref{model} the gauge fields, the action, and conjugate momenta are defined, from which Maxwell equations and magnetic charges arise. In Section \ref{T} the energy momentum tensor is computed, showing that the energy density is on-shell bounded from below. Finally Section \ref{wittenMax} shows that no Witten-like effect exists fo this model: magnetic charges are already intrinsically present in the theory. Having a well defined model for standard electromagnetism with magnetic charges, Section \ref{FractonDual} extends the idea to the covariant fracton theory and its self-duality. In Section \ref{FractonModel} the theory is defined, giving the self-dual Maxwell-like equations, and in Section \ref{matter} matter is introduced, thus showing the existence of fractonic magnetic charges. Section \ref{WittenFrac} shows that also in the fractonic case no Witten-like effect is present. Finally Section \ref{concl} summarize the results and draws conclusions, and Appendix \ref{appA} shows a particular situation in which a linearized gravity term is also taken into account.\\

\normalcolor

{\bf Notations and conventions :}\\
Minkowski metric $\eta_{\mu\nu}=\mbox{diag}(-1,1,1,1)$;\\
indices : $\mu,\nu,\rho,...=\{0,1,2,3\}$ and $i,j,k,...=\{1,2,3\}$;\\
Levi-Civita symbol $\epsilon_{0123}=1=-\epsilon^{0123}$, for which
	\begin{align}
	\epsilon^{0akp}\epsilon_{0biq}&=-\delta^a_b(\delta^k_i\delta^p_q-\delta^p_i\delta^k_q)+\delta^a_i(\delta^k_b\delta^p_q-\delta^p_b\delta^k_q)-\delta^a_q(\delta^k_b\delta^p_i-\delta^p_b\delta^k_i)\label{eps1}\\
	\epsilon^{0akp}\epsilon_{0aiq}&=-(\delta^k_i\delta^p_q-\delta^p_i\delta^k_q)\label{eps2}\ .
	\end{align}

\section{EM self-duality in generalized Maxwell theory}\label{EMduality}

\subsection{Maxwell equations, double potentials, and magnetic charges}\label{model}

In this Section, we present the EM self-duality \cite{Bunster:2011qp} in the case of generalized Maxwell theory that incorporates two independent gauge fields, one associated to the electric sector -$A_\mu(x)$- and the other one to the magnetic sector -$V_\mu(x)$-. Notice that in the context of self-dualities, \cite{Bunster:2011qp} discusses a different notion (dubbed ``twisted self-duality'' by the authors) compared to the one we consider in this manuscript. In particular, Bunster and Henneaux in \cite{Bunster:2011qp} define twisted self-duality by introducing two potentials (electric and magnetic one-forms $A_\mu(x)$ and $B_{\mu}(x)$), whose corresponding field strengths are dual to each other. This duality reduces the degrees of freedom of their action, making their theory equivalent to standard Maxwell theory. In contrast, the self-duality we will discuss refers to the invariance of our action under the transformation \eqref{EBdual2}, which involves the exchange of $A_\mu(x)$ and $V_\mu(x)$, as we shall see. In other words, our action will be dual to itself (namely self-dual). Crucially, our construction does not reduce the degrees of freedom: our two gauge fields $A_\mu(x)$ and $V_\mu(x)$ remain independent throughout, and thus our theory describes a generalization of the Maxwell theory with doubled degrees of freedom ($i.e.$ two independent gauge fields). With the aim of analysing the EM self-duality in the framework of an electromagnetic theory, we consider the following generalized Maxwell action 
	\be\label{Sem-dual}
		\begin{split}
		S_{em}&=\tfrac{\tilde a}{4}S_{max}[A]+\tfrac{\tilde b}{4}S_{max}[V]+\tfrac{\tilde \theta}{2}S_{\theta}[A,V]\ ,
		\end{split}
	\ee
where $\tilde{a}$, $\tilde{b}$ and $\tilde{\theta}$ are real parameters and
	\begin{align}
	S_{max}[A]&\equiv\int d^4x\,F^{\mu\nu}F_{\mu\nu}\\
	S_{max}[V]&\equiv\int d^4x\,G^{\mu\nu}G_{\mu\nu}\\
	S_{\theta}[A,V]&\equiv\int d^4x\,\epsilon^{\mu\nu\rho\sigma}F_{\mu\nu}G_{\rho\sigma}\ ,\label{SthetaAV}
	\end{align}
with
	\be
	F_{\mu\nu}\equiv \partial_\mu A_\nu-\partial_\nu A_\mu\quad;\quad G_{\mu\nu}\equiv \partial_\mu V_\nu-\partial_\nu V_\mu\ ,
	\ee
the field strengths associated to two gauge fields $A_\mu(x)$ and $V_\mu(x)$ which transform as follows
	\be
	\delta A_\mu=\partial_\mu\tilde\lambda\quad;\quad\delta V_\mu=\partial_\mu\tilde\lambda'\ ,
	\ee
where $\tilde\lambda(x)$ and $\tilde \lambda'(x)$ are local gauge parameters. Notice that the $S_{\theta}$ term \eqref{SthetaAV} is known as \textit{mutual theta term} and has been previously introduced in the context of topological phases of matter \cite{Palumbo:2013gxa,Ye:2013upa}. The action $S_{em}$ \eqref{Sem-dual}, given by the sum of (2.2)-\eqref{SthetaAV}, is postulated as a general action invariant under the gauge transformations (2.6), and containing Maxwell terms ((2.2) and (2.3)) and the mutual theta-term \eqref{SthetaAV}. The coefficients $\tilde a,\ \tilde b,\ \tilde\theta$ are \textit{a priori} general and kept free, in order to be tuned later to obtain the Maxwell equations and the self-duality property \eqref{EBdual2} we are looking for. When $\tilde{\theta}$ is constant, the mutual theta term contributes only as a boundary term, namely a three-dimensional BF term \cite{Palumbo2,Bertolini:2022sao} and does not affect the EoM, given by
	\begin{align}
	\frac{\delta S_{em}}{\delta A_{\alpha}}&=\tilde a\partial_\mu F^{\alpha\mu}\label{EoMAem}\\
	\frac{\delta S_{em}}{\delta V_{\alpha}}&=\tilde b\partial_\mu G^{\alpha\mu}\ .\label{EoMVem}
	\end{align}
However, it does modify the conjugate momenta as follows
	\begin{align}
	\Pi_{(\textsc{a})}^{a}\equiv\frac{\delta S_{em}}{\delta\partial_0A_{a}}&=-\tilde aF^{a0}+\tilde\theta\epsilon^{0amn}G_{mn}\label{PiA}\\
	\Pi_{(\textsc{v})}^{a}\equiv\frac{\delta S_{em}}{\delta\partial_0V_{a}}&=-\tilde bG^{a0}+\tilde\theta\epsilon^{0amn}F_{mn}\ .\label{PiV}
	\end{align}
Keeping in mind the following relations
	\begin{align}
	F^{a0}&=-\tfrac{1}{\tilde a}\Pi_{(\textsc{a})}^{a}+\tfrac{\tilde\theta}{\tilde a}\epsilon^{0amn}G_{mn}\\
	G^{a0}&=-\tfrac{1}{\tilde b}\Pi_{(\textsc{v})}^{a}+\tfrac{\tilde\theta}{\tilde b}\epsilon^{0amn}F_{mn}\\
	F_{mn}&=-\tfrac{1}{2\tilde\theta}\epsilon_{0amn}\left(\Pi_{(\textsc{v})}^{a}+\tilde bG^{a0}\right)\\
	G_{mn}&=-\tfrac{1}{2\tilde\theta}\epsilon_{0amn}\left(\Pi_{(\textsc{a})}^{a}+\tilde aF^{a0}\right)\ ,
	\end{align}
we can rewrite the EoM \eqref{EoMAem} and \eqref{EoMVem} in terms of the conjugate momenta \eqref{PiA} and \eqref{PiV}. From the EoM \eqref{EoMAem} of $A_\mu(x)$ we have
	\begin{align}
	\frac{\delta S_{em}}{\delta A_{0}}&=-\tilde a\partial_mF^{m0}=\partial_m\Pi_{(\textsc{a})}^{m}\label{EoMem1}\\
	\frac{\delta S_{em}}{\delta A_{a}}&=-\tilde a\partial_\mu F^{\mu a}=-\partial_0\Pi_{(\textsc{a})}^{a}+\left(2\tilde\theta+\tfrac{\tilde a\tilde b}{2\tilde\theta}\right)\epsilon^{0abc}\partial_0\partial_bV_c+\tfrac{\tilde a}{2\tilde\theta}\epsilon^{0abc}\partial_b\Pi_{(\textsc{v})c}\ ,
	\end{align}
and analogously, from the EoM \eqref{EoMVem} of $V_\mu(x)$ we have
	\begin{align}
	\frac{\delta S_{em}}{\delta V_{0}}&=-\tilde b\partial_mG^{m0}=\partial_m\Pi_{(\textsc{v})}^{m}\\
	\frac{\delta S_{em}}{\delta V_{a}}&=-\tilde b\partial_\mu G^{\mu a}=-\partial_0\Pi_{(\textsc{v})}^{a}+\left(2\tilde\theta+\tfrac{\tilde a\tilde b}{2\tilde\theta}\right)\epsilon^{0abc}\partial_0\partial_bA_c+\tfrac{\tilde b}{2\tilde\theta}\epsilon^{0abc}\partial_b\Pi_{(\textsc{a})c}\ .\label{EoMem4}
	\end{align}
There exists a special case for which all equations can be written only in terms of the conjugate momenta, that is when
	\be\label{constraint}
	\tilde a\tilde b=-4\tilde\theta^2\ .
	\ee
Thus defining
	\begin{align}
	E^a&\equiv\Pi_{(\textsc{a})}^{a}\\
	B^a&\equiv\Pi_{(\textsc{v})}^{a}\ ,
	\end{align}
the four EoM \eqref{EoMem1}-\eqref{EoMem4}, on-shell, become
	\begin{align}
	\partial_mE^{m}&=0\label{EoMem1}\\
	\partial_0E^{a}-\tfrac{\tilde a}{2\tilde\theta}\epsilon^{0abc}\partial_bB_c&=0\label{EoMem2}\\
	\partial_mB^{m}&=0\label{EoMem3}\\
	\partial_0B^{a}+\tfrac{2\tilde\theta}{\tilde a}\epsilon^{0abc}\partial_bE_c&=0\ .\label{EoMem4}
	\end{align}
By finally setting
	\be\label{condEM}
	\tilde a=-2\tilde\theta=-\tilde b\ ,
	\ee
we recover the standard Maxwell equations
	\begin{empheq}{align}
	\vec\nabla\cdot\vec E&=0\label{max1}\\
	\vec\nabla\cdot\vec B&=0\label{max2}\\
	\vec\nabla\times\vec E-\partial_t\vec B&=0\label{max3}\\
	\vec\nabla\times\vec B+\partial_t\vec E&=0\ .\label{max4}
	\end{empheq}
Therefore under the above condition \eqref{condEM}, the action \eqref{Sem-dual} that generates Maxwell equations is given by
	\be\label{Sem-dual'}
		\begin{split}
		S_{em}&=\tfrac{\tilde \theta}{2}\left(-S_{max}[A]+S_{max}[V]+S_{\theta}[A,V]\right)\\
		&=\tfrac{\tilde\theta}{2}\int d^4x\left(-F^{\mu\nu}F_{\mu\nu}+G^{\mu\nu}G_{\mu\nu}+\epsilon^{\mu\nu\rho\sigma}F_{\mu\nu}G_{\rho\sigma}\right)\ ,
		\end{split}
	\ee
whose conjugate momenta are the electric $and$ magnetic fields
	\begin{align}
	\Pi_{(\textsc{a})}^{a}&=E^a=2\tilde\theta \left( F^{a0}+\tfrac{1}{2}\epsilon^{0amn}G_{mn}\right)\label{E(F,G)}\\
	\Pi_{(\textsc{v})}^{a}&=B^a=2\tilde \theta\left(- G^{a0}+\tfrac{1}{2}\epsilon^{0amn}F_{mn}\right)\ .\label{B(F,G)}
	\end{align}
We thus recover equations that formally look like the standard Maxwell equations, although both fields $E_i(x)$ and $B_i(x)$ are constructed from our two independent gauge fields $A_\mu(x)$ and $V_\mu(x)$ through \eqref{E(F,G)} and \eqref{B(F,G)}. For this reason \eqref{max1}-\eqref{max4} represent the equations of motion of a generalized Maxwell action given by \eqref{Sem-dual'}.
Concerning the action, we can also see that it is still true that
	\begin{align}
	S_{em}&=\frac{1}{4\tilde\theta}\int d^4x\left(E_aE^a-B_aB^a\right)\label{E-B}\\
	&=\tilde\theta\int d^4x\left[\left(F^{a0}+\tfrac{1}{2}\epsilon^{0amn}G_{mn}\right)\left(-F_{a0}-\tfrac{1}{2}\epsilon_{0abc}G^{bc}\right)-\right.\nonumber\\
	&\left.\qquad\qquad-\left(-G^{a0}+\tfrac{1}{2}\epsilon^{0amn}F_{mn}\right)\left(G_{a0}-\tfrac{1}{2}\epsilon_{0abc}F^{bc}\right)\right]\nonumber\\
	&=\tfrac{\tilde\theta}{2}\int d^4x\left[-2F^{a0}F_{a0}-2\epsilon_{0abc}F^{a0}G^{bc}+G_{mn}G^{mn}+2G^{a0}G_{a0}-2\epsilon_{0abc}G^{a0}F^{bc}-F_{mn}F^{mn}\right]\nonumber\\
	&=\tfrac{\tilde\theta}{2}\int d^4x\left(-F^{\mu\nu}F_{\mu\nu}+G^{\mu\nu}G_{\mu\nu}+\epsilon^{\mu\nu\rho\sigma}F_{\mu\nu}G_{\rho\sigma}\right)\nonumber\ .
	\end{align}
A self-duality is therefore established for
	\be\label{AVdual}
	A_\mu\to-V_\mu\quad;\quad V_\mu\to A_\mu\ ,
	\ee
which implies the EM duality \eqref{EBdual2}.
 The action $S_{em}$ \eqref{Sem-dual'} under the transformation \eqref{AVdual} changes of an overall sign
	\be
	S_{em}\to S_{em}'=\tfrac{\tilde \theta}{2}\left(-S_{max}[V]+S_{max}[A]-S_{\theta}[A,V]\right)=-S_{em}\ ,
	\ee
which is not surprising, given that this is in agreement with how the action $S_{em}$ \eqref{E-B} in terms of the electromagnetic fields $E^a(x)$ and $B^a(x)$ transforms under this exchange of electric and magnetic fields \eqref{EBdual2}. 
We remind here that in general $\tilde{\theta}$ can be seen as a real and compact field, namely 
	\be
	\tilde{\theta}\rightarrow {\tilde{\theta}}+ 2\pi\ .
	\ee
Thus, to invert the sign flip of $S_{em}'$, we require the following transformation 
	\be
	\tilde \theta\ \to\ -\tilde\theta\ ,
	\ee
which implies that 
	\be
	\tilde{\theta}=\pi\ .
	\ee
This argument is similar to impose time-reversal invariance in Axion Electrodynamics in the context of three-dimensional topological insulators in class AII \cite{Qi:2008ew}. Concerning the generality of the choice \eqref{condEM}, notice that a redefinition of the fields is always possible, which does not affect the signs of the tuned doubled-Maxwell action \eqref{Sem-dual'} and would simply imply a change in the self-duality transformation \eqref{AVdual}. For instance one can redefine 
\be
A_\mu\to A'_\mu=aA_{\mu}\quad;\quad V_\mu\to V'_\mu=vV_{\mu}\ ,
\ee
with $a$ and $v$ constants, which would correspond to the following new self-dual  transformation
\be
A_\mu\to -\frac{v }{a}V_{\mu}\quad;\quad V_\mu\to \frac{a}{v}A_\mu\ .
\ee
The choice \eqref{condEM} simplify calculations and makes the self-dual invariance of Maxwell equations \eqref{max1}-\eqref{max4} and action \eqref{Sem-dual'} more evident. We can say that the choice \eqref{condEM} is not mandatory and it can be seen as a redefinition of the fields that make calculations and results more evident. On the other hand the choice \eqref{constraint} is a real tuning of our action. Indeed since we have two independent gauge fields, there is only one real coupling (out of the three $\tilde a,\ \tilde b,\ \tilde\theta$) to the action that cannot be reabsorbed by a redefinition of the fields, and that one is fixed by \eqref{constraint}. That tuning is the one necessary to have equations that can be written in terms of the electric and magnetic fields $E^a(x)$ and $B^a(x)$. Without it no Maxwell/electromagnetic interpretation can be given to the theory, and thus no other electromagnetic-like phenomena could be discussed at all. We can now introduce the matter sources into the action
	\be\label{Sem-dual'+source}
	S_{tot}=S_{em}+S_J+S_K=\int d^4x\left[\tfrac{\pi}{2}\left(-F^{\mu\nu}F_{\mu\nu}+G^{\mu\nu}G_{\mu\nu}+\epsilon^{\mu\nu\rho\sigma}F_{\mu\nu}G_{\rho\sigma}\right)-A_\mu \tilde J^\mu-V_\mu \tilde K^\mu\right]\ ,
	\ee
where
	\begin{align}
	S_J&=-\int d^4x\, A_\mu \tilde J^\mu\label{SJ}\\
	S_K&=-\int d^4x\, V_\mu\tilde  K^\mu\ ,\label{SK}
	\end{align}
the EoM are modified as follows
	\begin{align}
	\frac{\delta S_{tot}}{\delta A_{\alpha}}&=-2\pi \,\partial_\mu F^{\alpha\mu}-\tilde J^\alpha\label{EoMAJem}\\
	\frac{\delta S_{tot}}{\delta V_{\alpha}}&=2\pi \,\partial_\mu G^{\alpha\mu}-\tilde K^\alpha\ ,\label{EoMVKem}
	\end{align}
As a consequence the Maxwell equations \eqref{EoMem1}-\eqref{EoMem4} become
	\begin{align}
	\partial_aE^{a}&=\tilde \rho_{_{(e)}}\\
	\partial_0E^{a}+\epsilon^{0abc}\partial_bB_c&=-\tilde J^a\\
	\partial_aB^{a}&=\tilde \rho_{_{(m)}}\\
	\partial_0B^{a}-\epsilon^{0abc}\partial_bE_c&=-\tilde K^a\ ,
	\end{align}
with 
	\be
	\tilde \rho_{_{(e)}}\equiv \tilde J^0\quad;\quad 	\tilde \rho_{_{(m)}}\equiv \tilde K^0\ .
	\ee
The self-duality in the EoM is here preserved if we consider the following transformations on the source terms
	\be\label{J->K}
	\tilde J^\alpha\ \to\ -\tilde K^\alpha\quad;\quad \tilde K^\alpha\ \to\ \tilde J^\alpha\ .
	\ee
A final remark is here required in order to better highlight the results. Through \eqref{constraint} and \eqref{condEM}, we obtained the tuned Maxwell-like action \eqref{Sem-dual'}. The question is now whether that tuned action \eqref{Sem-dual'} is equivalent to a Maxwell one, and what exactly means ``equivalent''. From the point of view of the EoM \eqref{max1}-\eqref{max4} they are formally Maxwell equations, $i.e.$ written in terms of gauge invariant (thus physical) quantities which can be identified as electromagnetic fields, both of which, in this case, are conjugate momenta of the theory. That is thanks to the presence of the mutual theta term \eqref{SthetaAV}, which thus enhance the duality of the theory. It is also interesting to notice that the tuned action \eqref{Sem-dual'} can be rewritten as a proper Maxwell term if we define a generalized field strength
\be
\Phi^{\mu\nu}\equiv F^{\mu\nu}-\tfrac{1}{2}\epsilon^{\mu\nu\rho\sigma}G_{\rho\sigma}\ ,
\ee
for which the action in \eqref{Sem-dual'} is
\be\label{SemPhi}
S_{em}=-\tfrac{\tilde\theta}{2}\int d^4x \Phi^{\mu\nu}\Phi_{\mu\nu}\ ,
\ee
and the electromagnetic fields $E_i(x)$ and $B_i(x)$ in \eqref{E(F,G)} and \eqref{B(F,G)} are
\be
E_a\propto\Phi_{a0}\quad;\quad B_a\propto\Phi^*_{a0}=\tfrac1 2 \epsilon_{0aij}\Phi^{ij}\ .
\ee
However we have to keep in mind the true ``doubled'' nature of the theory, which relies on two gauge fields and emerges when matter sources $\tilde J_\mu(x)$ and $\tilde K_\mu(x)$ are introduced: it has twice the number of degrees of freedom, and, most importantly, allows for the presence of magnetic charges. Therefore the tuned action $S_{em}$ \eqref{Sem-dual'} describes a generalized Maxwell theory, as it reproduces most of its features, but it is not equivalent to the standard one, since it allows, through $S_{tot}$ \eqref{Sem-dual'+source}, for the presence of an additional physical quantity: the magnetic charge

\subsection{Energy-momentum tensor and energy density}\label{T}
We want now to study the behaviour of the energy density of our theory. The energy-momentum tensor is computed to be
	\be
		\begin{split}
		T_{\alpha\beta}=&-\frac{2}{\sqrt{-g}}\frac{\delta S_{em}}{\delta g^{\alpha\beta}}\\
		=&-\frac{\tilde\theta}{\sqrt{-g}}\frac{\delta }{\delta g^{\alpha\beta}}\int d^4x\sqrt{-g}g^{\mu\rho}g^{\nu\sigma}\left(-F_{\mu\nu}F_{\rho\sigma}+G_{\mu\nu}G_{\rho\sigma}\right)\\
		=&\tilde\theta\left[\frac{1}{2}g_{\alpha\beta}\left(-F_{\mu\nu}F^{\mu\nu}+G_{\mu\nu}G^{\mu\nu}\right)+2\left(F_{\alpha\mu}F^{\ \mu}_\beta-G_{\alpha\mu}G^{\ \mu}_\beta\right)\right]\ ,
		\end{split}
	\ee
which represents two copies of a Maxwell energy-momentum tensor and, as the standard one, it is conserved on-shell, $i.e.$
	\be
	\partial^\beta T_{\alpha\beta}=0\ .
	\ee
Its 00-component, $i.e.$ the energy density, is not evidently positive definite. Indeed
	\be
		\begin{split}
		T_{00}=&\frac{\tilde\theta}{2}\left(-2F_{m0}F^{m0}+F_{mn}F^{mn}+2G_{m0}G^{m0}-G_{mn}G^{mn}\right)\\
		=&\frac{1}{4\tilde\theta}\left(E_mE^m-B_mB^m\right)+\frac{1}{2}\epsilon^{0mnp}\left(B_mF_{np}-E_mG_{np}\right)\\
		=&\frac{1}{4\tilde\theta}\left(E_mE^m+B_mB^m\right)-\frac{1}{2}\epsilon^{0mnp}E_mG_{np}+B_mG^{m0}\ .
		\end{split}
	\ee
The issue was for instance addressed in \cite{CasteloFerreira:2005ej} for a similar model, proposing various possibilities for overcoming it. However one can notice that 
	\be
	T_{00}=\frac{1}{4\tilde\theta}\left(E_mE^m+B_mB^m\right)+V_a\left(\partial_0B^a-\epsilon^{0amn}\partial_mE_n\right)-V_0\partial_mB^m+\partial T
	\ee
where ``$\partial T$'' are total derivative terms, in particular
	\be
	\partial T\equiv\partial_m(\epsilon^{0amn}V_aE_n+V_0B^m)-\partial_0(V_mB^m)\ .
	\ee
Therefore on-shell, $i.e.$ using the EoM \eqref{EoMem3} and \eqref{EoMem4}, up to boundary terms, we have
	\be
	T_{00}=\frac{1}{4\tilde\theta}\left(E_mE^m+B_mB^m\right)\ ,
	\ee
which is positive-definite when
	\be
	\tilde\theta>0\ .
	\ee
Notice that the sign in front of \eqref{SemPhi}, which corresponds to the usual sign in front of Maxwell action for positive-definite energy, confirms the above constraint  $\tilde\theta>0$. Thus, our case with $\tilde\theta=\pi$ has an energy density bounded from below.

\subsection{Axion-like terms: no Witten effect}\label{wittenMax}
One of the main non-perturbative features of Axion Electrodynamics concerns the existence of the Witten effect \cite{Witten:1979ey,Rosenberg:2010ia}, for which a magnetic monopole acquires an electric charge and becomes a dyon, namely a new particle that carries both electric and magnetic charges. In order to study a possible Witten-like effect in our theory, we would need to promote the mutual theta-term $S_{\theta}$ \eqref{SthetaAV} to have a spacetime dependence on $\theta=\theta(x)$. However, this implies a modification of the original conjugate momenta of the theory
	\begin{align}
	\frac{\delta S_{em}}{\delta\partial_0A_{a}}&=E^{a}=-\tilde aF^{a0}+\tilde\theta(x)\epsilon^{0amn}G_{mn}\\
	\frac{\delta S_{em}}{\delta\partial_0V_{a}}&=B^{a}=-\tilde bG^{a0}+\tilde\theta(x)\epsilon^{0amn}F_{mn}\ .
\end{align}
We instead decide to keep the canonical momenta given by \eqref{PiA} and \eqref{PiV}, $i.e.$ those with a constant $\tilde{\theta}$, and add to the action $S_{em}$ \eqref{Sem-dual} axion-like contributions $S_{\tilde\theta_1}$ and $S_{\tilde\theta_2}$ that do not mix the two gauge sectors. The new action is defined by
	\begin{align}
	S_{tot}&=S_{em}+S_{\tilde\theta_1}+S_{\tilde\theta_2}\label{Sem-dual-theta2}\ ,
	\end{align}
with
	\begin{align}
	S_{\tilde\theta_1}&=\frac{1}{4}\int d^4x\,\tilde\theta_1\epsilon^{\mu\nu\rho\sigma}F_{\mu\nu}F_{\rho\sigma}\\
	S_{\tilde\theta_2}&=\frac{1}{4}\int d^4x\,\tilde\theta_2\epsilon^{\mu\nu\rho\sigma}G_{\mu\nu}G_{\rho\sigma}\ ,
	\end{align}
and
	\be
	\tilde\theta={\rm {const.}}\quad;\quad\tilde\theta_1=\tilde\theta_1(x)\quad;\quad\tilde\theta_2=\tilde\theta_2(x)\ .
	\ee
 The corresponding EoM are given by
	\begin{align}
	\frac{\delta S_{tot}}{\delta A_{\alpha}}&=\tilde a\partial_\mu F^{\alpha\mu}+\epsilon^{\alpha\mu\nu\rho}\partial_\mu\tilde\theta_1\, F_{\nu\rho}\label{EoMAem-theta2}\\
	\frac{\delta S_{tot}}{\delta V_{\alpha}}&=\tilde b\partial_\mu G^{\alpha\mu}+\epsilon^{\alpha\mu\nu\rho}\partial_\mu\tilde\theta_2\, G_{\nu\rho}\ .\label{EoMVem-theta2}
	\end{align}
The components of the EoM translates as follows
	\bi
	\item from \eqref{EoMAem-theta2} we have
	\begin{align}
	\frac{\delta S_{tot}}{\delta A_{0}}=&\partial_mE^{m}+\frac{1}{\tilde\theta}\partial_m\tilde\theta_1\left(B^m+\tilde b G^{m0}\right)\label{EoMem1-theta2}\\
	\frac{\delta S_{tot}}{\delta A_{a}}=&-\partial_0E^{a}+\tfrac{\tilde a}{2\tilde\theta}\epsilon^{0abc}\partial_bB_c-\frac{1}{\tilde\theta}\dot{\tilde\theta}_1(B^a+\tilde b G^{a0})-\frac{2}{\tilde a}\epsilon^{0amn}\partial_m\tilde\theta_1\left(E_n+\tilde\theta\epsilon_{0nbc}G^{bc}\right)\ ,
	\end{align}
where the constraint \eqref{constraint} on the parameter has been used.
	\item From \eqref{EoMVem-theta2} we have
	\begin{align}
	\frac{\delta S_{tot}}{\delta V_{0}}=&\partial_mB^{m}+\frac{1}{\tilde\theta}\partial_m\tilde\theta_2\left(E^m+\tilde a F^{m0}\right)\\
	\frac{\delta S_{tot}}{\delta V_{a}}=&-\partial_0B^{a}+\tfrac{\tilde b}{2\tilde\theta}\epsilon^{0abc}\partial_bE_c-\frac{1}{\tilde\theta}\dot{\tilde\theta}_2\left(E^a+\tilde a F^{a0}\right)-\frac{2}{\tilde b}\epsilon^{0amn}\partial_m\tilde\theta_2\left(B_n+\tilde\theta\epsilon_{0nbc}F^{bc}\right)\ .
	\end{align}
	\ei
Therefore setting
	\be\label{condEM2}
	\tilde a=-2\tilde\theta=-\tilde b\ ,
	\ee
we get the following on-shell equations
	\begin{align}
	\partial_mE^{m}&=-\frac{1}{\tilde\theta}\partial_m\tilde\theta_1\left(B^m+2\tilde \theta G^{m0}\right)\label{EoMem1-theta2'}\\
	\partial_0E^{a}+\epsilon^{0abc}\partial_bB_c&=\frac{1}{\tilde\theta}\left[-\dot{\tilde\theta}_1\left(B^a+2\tilde \theta G^{a0}\right)+\epsilon^{0amn}\partial_m\tilde\theta_1\left(E_n+\tilde\theta\epsilon_{0nbc}G^{bc}\right)\right]\\
	\partial_mB^{m}&=-\frac{1}{\tilde\theta}\partial_m\tilde\theta_2\left(E^m-2\tilde \theta F^{m0}\right)\\
	-\partial_0B^{a}+\epsilon^{0abc}\partial_bE_c&=\frac{1}{\tilde\theta}\left[\dot{\tilde\theta}_2\left(E^a-2\tilde \theta F^{a0}\right)+\epsilon^{0amn}\partial_m\tilde\theta_2\left(B_n+\tilde\theta\epsilon_{0nbc}F^{bc}\right)\right]\ .\label{EoMem4-theta2'}
	\end{align}
However we cannot write the EoM only in terms of the conjugate momenta. Indeed, as a consequence of our doubled-potential approach, there appear four possible vector invariants of the theory, namely
	\be
	X_1^a\equiv F^{a0}\quad;\quad X_2^a\equiv\epsilon^{0abc}F_{bc}\quad;\quad X_3^a\equiv G^{a0}\quad;\quad X_4^a\equiv\epsilon^{0abc}G_{bc}\ .
	\ee
Invariants typically represent physical quantities and in this case two combinations give the canonical momenta, $i.e.$ the electric and magnetic fields
	\be\label{E-B-witten}
	E^a=\tilde\theta\left(2X_1^a+X_4^a\right)\quad;\quad 	B^a=\tilde\theta\left(-2X_3^a+X_2^a\right)\ ,
	\ee
and the remaining two, $X_1^a(x)$ and $X_3^a(x)$, get involved in the EoM when axion-like terms are introduced, and interplay with the electric and magnetic fields as effective charges and currents. In particular the EoM \eqref{EoMem1-theta2'}-\eqref{EoMem4-theta2'} can be written as
	\begin{align}
	\partial_mE^{m}&=-\partial_m\tilde\theta_1\left(\tfrac{1}{\tilde\theta}B^m+2 X^m_3\right)\label{EoM1axion}\\
	\partial_0E^{a}+\epsilon^{0abc}\partial_bB_c&= -\dot{\tilde\theta}_1\left(\tfrac{1}{\tilde\theta}B^a+2 X^a_3\right)+2\epsilon^{0amn}\partial_m\tilde\theta_1 X_{1\,n} \\
	\partial_mB^{m}&=-\partial_m\tilde\theta_2\left(\tfrac{1}{\tilde\theta}E^m-2 X_1^m\right)\\
	-\partial_0B^{a}+\epsilon^{0abc}\partial_bE_c&= \dot{\tilde\theta}_2\left(\tfrac{1}{\tilde\theta}E^a-2X_1^a\right)-2\epsilon^{0amn}\partial_m\tilde\theta_2 X_{3\,n} \ .\label{EoM4axion}
	\end{align}
We can however see that the presence of these additional contributions $X_1(x)$ and $X_3(x)$ prevent the existence of a Witten-like effect. Indeed, by considering for simplicity only time-dependence on the $\tilde\theta_{1,2}(x)=\tilde\theta_{1,2}(t)$ \cite{Rosenberg:2010ia}, the above EoM \eqref{EoM1axion}-\eqref{EoM4axion} coupled to matter through the terms $S_J$ \eqref{SJ} and $S_K$ \eqref{SK}, become
	\begin{align}
	\partial_mE^{m}&=\tilde \rho_{_{(e)}}\label{EoM1axionT}\\
	\partial_0E^{a}+\epsilon^{0abc}\partial_bB_c&= -\dot{\tilde\theta}_1\left(\tfrac{1}{\tilde\theta}B^a+2 X^a_3\right)-\tilde J^a\label{EoM2axionT} \\
	\partial_mB^{m}&=\tilde \rho_{_{(m)}}\\
	\partial_0B^{a}-\epsilon^{0abc}\partial_bE_c&= -\dot{\tilde\theta}_2\left(\tfrac{1}{\tilde\theta}E^a-2X_1^a\right)-\tilde K^a \ .\label{EoM4axionT}
	\end{align}
Typically, in the non-self-dual case, taking the spatial divergence $\partial_a$ of the modified Amp\`ere equation and using the electric Gauss law, relates the divergence of the magnetic field to the electric charge, through the additional theta-term. In this case, however, if we take the spatial divergence $\partial_a$ of the ``dual'' Amp\`re equation \eqref{EoM2axionT}, the $\tilde\theta_1$-contribution vanishes because
	\be\label{X3=B}
	\partial_a X_3^a=-\frac{1}{2\tilde\theta}\partial_aB^a\ ,
	\ee 
an we are left with the standard electric continuity equation
	\be
	\partial_0\tilde \rho_{_{(e)}}+\partial_a\tilde J^a=0\ .
	\ee
The same happens with the second set of equations, for which we only get the magnetic continuity equation
	\be
	\partial_0\tilde \rho_{_{(m)}}+\partial_a\tilde K^a=0\ .
	\ee
Thus no Witten-like effect ($i.e.$ no dyonic charges) is present in the self-dual model. This was somehow expected, since the self-dual model already provides both an electric and a magnetic charges, associated to two independent gauge fields. As already remarked, the choice \eqref{condEM} simplify calculations and makes the self-dual invariance of Maxwell equations \eqref{max1}-\eqref{max4} and action \eqref{Sem-dual'} more evident. It does not, however, affect the physics, such as the absence of the Witten effect, since it is only a redefinition of the fields. In fact, not considering that choice (or another combination) would correspond to a different numerical factor in the definition of magnetic field $B^a(x)$ in \eqref{E-B-witten}, and thus a different coefficient in \eqref{X3=B}, for which \eqref{EoM1axion} would change on the r.h.s., in such a way that the Witten effect would still be absent.

\section{Fractonic self-duality}\label{FractonDual}

\subsection{Covariant fractons with two tensor gauge fields}\label{FractonModel}
We are now ready to discuss the fractonic self-duality by generalizing the results presented in the previous sections to the case of covariant fractons. 
Thus, we consider two rank-2 symmetric gauge fields $A_{\mu\nu}(x)$ and $V_{\mu\nu}(x)$ which transform under the covariant fracton symmetry of \cite{Blasi:2022mbl,Bertolini:2022ijb,Bertolini:2023juh,Bertolini:2023sqa,Bertolini:2024yur,Afxonidis:2023pdq}
	\be\label{dfract}
	\delta A_{\mu\nu}=\partial_\mu\partial_\nu\lambda\quad;\quad\delta' V_{\mu\nu}=\partial_\mu\partial_\nu\lambda'\ .
	\ee
Invariant field strengths can be defined \cite{Bertolini:2022ijb}
	\begin{align}
	F_{\mu\nu\rho}&=F_{\nu\mu\rho}\equiv\partial_\mu A_{\nu\rho}+\partial_\nu A_{\rho\mu}-2\partial_\rho A_{\mu\nu}\label{defF}\\
	G_{\mu\nu\rho}&=G_{\nu\mu\rho}\equiv\partial_\mu V_{\nu\rho}+\partial_\nu V_{\rho\mu}-2\partial_\rho V_{\mu\nu}\ ,\label{defG}
	\end{align}
$i.e.$ such that
	\be
	\delta F_{\mu\nu\rho}=0\quad;\quad \delta'G_{\mu\nu\rho}=0\ ,
	\ee
which also imply the following cyclicity property
	\be\label{cicl}
	F_{\mu\nu\rho}+F_{\nu\rho\mu}+F_{\rho\mu\nu}=0\quad;\quad G_{\mu\nu\rho}+G_{\nu\rho\mu}+G_{\rho\mu\nu}=0\ ,
	\ee
and corresponding Bianchi-like identities
	\be\label{bianchi}
	\epsilon_{\alpha\mu\nu\rho}\partial^{\mu}F^{\beta\nu\rho}=0\quad;\quad \epsilon_{\alpha\mu\nu\rho}\partial^{\mu}G^{\beta\nu\rho}=0\ .
	\ee
The corresponding gauge-invariant fractonic action is given by
\be\label{Sfracton-dual}
S=\int d^4x\left(\tfrac{a}{6}F^{\mu\nu\rho}F_{\mu\nu\rho}+\tfrac{b}{6}G^{\mu\nu\rho}G_{\mu\nu\rho}+\tfrac{2}{3}\theta\epsilon^{\mu\nu\rho\sigma}F^\lambda_{\ \mu\nu}G_{\lambda\rho\sigma}\right)\ ,
\ee
where two of the three constants can be reabsorbed, up to a sign, by a redefinition of the fields, and where the gauge fields have the following mass dimensions
	\be
	[A_{\mu\nu}]=[V_{\mu\nu}]=1\ .
	\ee
The last term in the action $S$ \eqref{Sfracton-dual}  represents a mutual $\theta$-like term  for fractons \cite{Bertolini:2022ijb}  and it is a pure boundary term. The other contributions to the action $S$ are an extension of the fractonic one studied in \cite{Bertolini:2022ijb}, when an additional gauge field, $V_{\mu\nu}$, that transform under the covariant fracton symmetry \eqref{dfract}, is introduced in the theory. We are interested in a self-duality similar to the electromagnetic one of Section \ref{EMduality}, and therefore in the introduction of magnetic charges for covariant fractons too. In the absence of a boundary, the $\theta$-term only affects the definition of the conjugate momenta, but not the EoM of the theory, which are
	\begin{align}
	\frac{\delta S}{\delta A_{\alpha\beta}}&=a\partial_\mu F^{\alpha\beta\mu}\label{EoMA}\\
	\frac{\delta S}{\delta V_{\alpha\beta}}&=b\partial_\mu G^{\alpha\beta\mu}\ .\label{EoMV}
	\end{align}
Instead, concerning the conjugate momenta we have
	\begin{align}
	\frac{\delta S}{\delta\partial_0A_{\alpha\beta}}&=\Pi_{(\textsc{a})}^{\alpha\beta}=-aF^{\alpha\beta0}+\theta\left(\epsilon^{0\alpha mn}G^\beta_{\ mn}+\epsilon^{0\beta mn}G^\alpha_{\ mn}\right)\\
	\frac{\delta S}{\delta\partial_0V_{\alpha\beta}}&=\Pi_{(\textsc{v})}^{\alpha\beta}=-bG^{\alpha\beta0}+\theta\left(\epsilon^{0\alpha mn}F^\beta_{\ mn}+\epsilon^{0\beta mn}F^\alpha_{\ mn}\right)\ ,
	\end{align}
thus
	\begin{align}
	\Pi_{(\textsc{a})}^{00}&=\Pi_{(\textsc{v})}^{00}=0\\
	\Pi_{(\textsc{a})}^{a0}&=-aF^{a00}-\theta\epsilon^{0amn}G_{0mn}\\
	\Pi_{(\textsc{v})}^{a0}&=-bG^{a00}-\theta\epsilon^{0amn}F_{0mn}\\
	\Pi_{(\textsc{a})}^{ab}&=-aF^{ab0}+\theta\left(\epsilon^{0amn}G^b_{\ mn}+\epsilon^{0bmn}G^a_{\ mn}\right)\label{fracPa}\\
	\Pi_{(\textsc{v})}^{ab}&=-bG^{ab0}+\theta\left(\epsilon^{0amn}F^b_{\ mn}+\epsilon^{0bmn}F^a_{\ mn}\right)\ .\label{fracPv}
	\end{align}
We now study the components of the EoM
	\bi
	\item $\pmb{\alpha=\beta=0}$ for \eqref{EoMA} and \eqref{EoMV} respectively 
		\begin{align}
		&a\partial_\mu F^{00\mu}=a\partial_mF^{00m}=2a\partial_m\left(\partial^0A^{0m}-\partial^mA^{00}\right)=2\partial_m\Pi_{(\textsc{a})}^m\label{EoMA00}\\
		&b\partial_\mu G^{00\mu}=b\partial_mG^{00m}=2b\partial_m\left(\partial^0V^{0m}-\partial^mV^{00}\right)=2\partial_m\Pi_{(\textsc{v})}^m\ ,\label{EoMV00}
		\end{align}
due to the cyclicity properties \eqref{cicl} and Bianchi identities \eqref{bianchi} and to the definitions of the conjugate momenta $\Pi_{(\textsc{a})}^{ab}$ \eqref{fracPa} and $\Pi_{(\textsc{v})}^{ab}$ \eqref{fracPv}. In the non-dual theory studied in \cite{Bertolini:2022ijb} it has been shown that the scalar charge theory for fractons \cite{Pretko:2016lgv,Pretko:2016kxt,Pretko:2017xar,Pretko:2020cko} is embedded in the covariant model through a particular solution of the on-shell EoM \eqref{EoMA00}, which is
	\begin{equation}\label{A0}
	A_{0\mu}=A_{\mu0}\equiv\partial_\mu A_0\ .
	\end{equation}
For the dual theory \eqref{Sfracton-dual}, this extends analogously to a particular solution of the EoM \eqref{EoMV00} as
	\be\label{V0}
	\quad V_{0\mu}=V_{\mu0}\equiv\partial_\mu V_0\ ,
	\ee
which we shall adopt from now on. These solutions are such that
	\begin{empheq}{align}
	&F^{i00}=F^{0i0}=F^{00i}=0\label{F00i=0}\\
	&G^{i00}=G^{0i0}=G^{00i}=0\label{G00i=0}\\
	&F^{ij0}=-2F^{0ij}=-2F^{i0j}\label{Fij0=-2F0ij}\\
	&G^{ij0}=-2G^{0ij}=-2G^{i0j}\label{Gij0=-2G0ij}\ ,
	\end{empheq}
and
	\begin{align}
	\Pi_{(\textsc{a})}^{00}&=\Pi_{(\textsc{v})}^{00}=\Pi_{(\textsc{a})}^{a0}=\Pi_{(\textsc{v})}^{a0}=0\\
	\Pi_{(\textsc{a})}^{ab}&=-aF^{ab0}+\theta\left(\epsilon^{0amn}G^b_{\ mn}+\epsilon^{0bmn}G^a_{\ mn}\right)\label{fracPaSol}\\
	\Pi_{(\textsc{v})}^{ab}&=-bG^{ab0}+\theta\left(\epsilon^{0amn}F^b_{\ mn}+\epsilon^{0bmn}F^a_{\ mn}\right)\ .\label{fracPvSol}
	\end{align}
	\item $\pmb{\alpha=a\ ,\ \beta=0}$ for \eqref{EoMA} and \eqref{EoMV} respectively 
		\begin{align}
		&a\partial_\mu F^{a0\mu}=a\partial_0F^{a00}+a\partial_mF^{a0m}=-\frac{a}{2}\partial_mF^{am0}=\frac{1}{2}\partial_m\Pi^{am}_{(\textsc a)}-\frac{\theta}{2}\epsilon^{0abc}\partial_mG^m_{\ bc}\\
		&b\partial_\mu G^{a0\mu}=b\partial_0G^{a00}+b\partial_mG^{a0m}=-\frac{b}{2}\partial_mG^{am0}=\frac{1}{2}\partial_m\Pi^{am}_{(\textsc v)}-\frac{\theta}{2}\epsilon^{0abc}\partial_mF^m_{\ bc}\ ,
		\end{align}
due to the solutions \eqref{A0} and \eqref{V0}, the Bianchi identities \eqref{bianchi} and definitions of the conjugate momenta $\Pi_{(\textsc{a})}^{ab}$ \eqref{fracPaSol} and $\Pi_{(\textsc{v})}^{ab}$ \eqref{fracPvSol}. If we now take the divergence $\partial_a$ of both EoM and go on-shell, we get
	\begin{align}
	\frac{1}{2}\partial_a\partial_b\Pi^{ab}_{(\textsc a)}&=0\\
	\frac{1}{2}\partial_a\partial_b\Pi^{ab}_{(\textsc v)}&=0\ ,
	\end{align}
which are Gauss-like constraints typical of fracton models \cite{Pretko:2016lgv,Pretko:2016kxt,Pretko:2017xar,Pretko:2020cko,Bertolini:2022ijb}.
	\item $\pmb{\alpha=a\ ,\ \beta=b}$ for \eqref{EoMA} and \eqref{EoMV} respectively give
		\begin{align}
		a\partial_\mu F^{ab\mu}&=a\partial_0F^{ab0}+a\partial_mF^{abm}\\
		&=-\partial_0\Pi^{ab}_{(\textsc a)}-\tfrac{3\theta}{2b}\left(\epsilon^{0amn}\partial_m\Pi^b_{(\textsc v)\,n}+\epsilon^{0bmn}\partial_m\Pi^a_{(\textsc v)\,n}\right)+\nonumber\\
		&\quad+\left(a+\tfrac{9\theta^2}{b}\right)\partial_mF^{abm}-\frac{9\theta^2}{2b}\left(\eta^{ab}\partial_mF_n^{\ nm}-\tfrac{1}{2}\partial^aF_m^{\ mb}-\tfrac{1}{2}\partial^bF_m^{\ ma}\right)\nonumber\\
		b\partial_\mu G^{ab\mu}&=b\partial_0G^{ab0}+b\partial_mG^{abm}\\
		&=-\partial_0\Pi^{ab}_{(\textsc v)}-\tfrac{3\theta}{2a}\left(\epsilon^{0amn}\partial_m\Pi^b_{(\textsc a)\,n}+\epsilon^{0bmn}\partial_m\Pi^a_{(\textsc a)\,n}\right)+\nonumber\\
		&\quad+\left(b+\tfrac{9\theta^2}{a}\right)\partial_mG^{abm}-\frac{9\theta^2}{2b}\left(\eta^{ab}\partial_mG_n^{\ nm}-\tfrac{1}{2}\partial^aG_m^{\ mb}-\tfrac{1}{2}\partial^bG_m^{\ ma}\right)\nonumber
		\end{align}
Thus if we take
		\be
		b=-\frac{9\theta^2}{a}\ ,
		\ee
we get the following fractonic equations with intrinsic currents
		\begin{align}
		-\partial_0 E^{ab}+\tfrac{1}{2}\tfrac{a}{3\theta}\left(\epsilon^{0amn}\partial_mB^b_{\,n}+\epsilon^{0bmn}\partial_mB^a_{\,n}\right)&=j_{(e)}^{ab}\label{fracMaxIntr1}\\
		-\partial_0B^{ab}-\tfrac{1}{2}\tfrac{3\theta}{a}\left(\epsilon^{0amn}\partial_mE^b_{\,n}+\epsilon^{0bmn}\partial_mE^a_{\,n}\right)&=j_{(m)}^{ab}	\ ,\label{fracMaxIntr2}
		\end{align}
where we defined
		\begin{align}
		E^{ab}&\equiv\Pi^{ab}_{(\textsc a)}\label{defE}\\
		B^{ab}&\equiv\Pi^{ab}_{(\textsc v)}\label{defB}\\
		j_{(e)}^{ab}&\equiv-\frac{a}{2}\left(\eta^{ab}\partial_mF_n^{\ nm}-\tfrac{1}{2}\partial^aF_m^{\ mb}-\tfrac{1}{2}\partial^bF_m^{\ ma}\right)\label{je}\\
		j_{(m)}^{ab}&\equiv \frac{9\theta^2}{2a}\left(\eta^{ab}\partial_mG_n^{\ nm}-\tfrac{1}{2}\partial^aG_m^{\ mb}-\tfrac{1}{2}\partial^bG_m^{\ ma}\right)\ .\label{jm}
		\end{align}
	\ei
Notice that this intrinsic current closely resembles the Einstein tensor $\mathcal G^{\mu\nu}(x)$ of linearized gravity (LG) when written in terms of the field strengths $F_{\mu\nu\rho}(x)$ \eqref{defF} and $G_{\mu\nu\rho}(x)$ \eqref{defG} \cite{Bertolini:2023wie}, $i.e.$
	\begin{align}
	\mathcal G^{ab}_{(e)}&\propto -\partial_cF^{abc}+\eta^{ab}\partial_mF_n^{\ nm}-\tfrac{1}{2}\partial^aF_m^{\ mb}-\tfrac{1}{2}\partial^bF_m^{\ ma}=-\partial_cF^{abc}-\frac{2}{a}j_{(e)}^{ab}\\
	\mathcal G^{ab}_{(m)}&\propto -\partial_cG^{abc}+\eta^{ab}\partial_mG_n^{\ nm}-\tfrac{1}{2}\partial^aG_m^{\ mb}-\tfrac{1}{2}\partial^bG_m^{\ ma}=-\partial_cG^{abc}+\frac{2a}{9\theta^2}j_{(m)}^{ab}\ ,
	\end{align}
and, additionally, the following relations hold
		\be\label{j=0}
		\partial_a\partial_bj^{ab}_{(e,m)}=\partial_a\partial_b\mathcal G^{ab}_{(e,m)}=0
		\ee
identically. In Appendix \ref{appA} we show that this current can disappear if a LG-like term is considered in the action $S$ \eqref{Sfracton-dual} with properly tuned parameters. However that would require the fractonic solutions \eqref{A0} and \eqref{V0} to be imposed, and not recovered.

\subsection{Matter coupling}\label{matter}
We can now consider the theory coupled to matter sources
	\be
	S_J=-\int d^4x\left(A_{\mu\nu}J^{\mu\nu}+V_{\mu\nu}K^{\mu\nu}\right)\ ,
	\ee
where
	\be
	J^{\mu\nu}=J^{\nu\mu}\quad;\quad K^{\mu\nu}=K^{\nu\mu}
	\ee
are rank-2 symmetric tensors coupled to the gauge fields. The total action is thus
	\be
	S_{tot}=S+S_J=\int d^4x \left(\tfrac{a}{6}F^{\mu\nu\rho}F_{\mu\nu\rho}-\tfrac{3\theta^2}{2a}G^{\mu\nu\rho}G_{\mu\nu\rho}+\tfrac{2}{3}\theta\epsilon^{\mu\nu\rho\sigma}F^\lambda_{\ \mu\nu}G_{\lambda\rho\sigma}-A_{\mu\nu}J^{\mu\nu}-V_{\mu\nu}K^{\mu\nu}\right) \ ,
	\ee
whose EoM are modified as
	\begin{align}
	\frac{\delta S_{tot}}{\delta A_{\alpha\beta}}&=a\partial_\mu F^{\alpha\beta\mu}-J^{\alpha\beta}\label{EoMAJ}\\
	\frac{\delta S_{tot}}{\delta V_{\alpha\beta}}&=-\tfrac{9\theta^2}{a}\partial_\mu G^{\alpha\beta\mu}-K^{\alpha\beta}\ .\label{EoMVK}
	\end{align}
Considering the on-shell particular solutions \eqref{A0} and \eqref{V0} which embeds standard fractons, the components of \eqref{EoMAJ} and \eqref{EoMVK} gives
	\bi
	\item $\pmb{\alpha=\beta=0}$ 
		\be
		J^{00}=K^{00}=0\ .
		\ee
	\item $\pmb{\alpha=a\ ,\ \beta=0}$ 
		\begin{align}
		\frac{1}{2}\partial_mE^{am}-\frac{\theta}{2}\epsilon^{0abc}\partial_mG^m_{\ bc}&=J^{a0}\\
		\frac{1}{2}\partial_mB^{am}-\frac{\theta}{2}\epsilon^{0abc}\partial_mF^m_{\ bc}&=K^{a0}\ ,
		\end{align}
	which, taking the divergence $\partial_a$ of both equations, lead to
		\begin{align}
		\partial_a\partial_bE^{ab}&=\rho_{(e)}\label{eGauss}\\
		\partial_a\partial_bB^{ab}&=\rho_{(m)}\ ,\label{mGauss}
		\end{align}
	where we have defined the fractonic electric and magnetic -like charges as
		\be
		\rho_{(e)}\equiv2\partial_aJ^{a0}\quad;\quad\rho_{(m)}\equiv2\partial_aK^{a0}\ .
		\ee
	\item $\pmb{\alpha=a\ ,\ \beta=b}$ finally give
		\begin{align}
		-\partial_0 E^{ab}+\tfrac{1}{2}\tfrac{a}{3\theta}\left(\epsilon^{0amn}\partial_mB^b_{\,n}+\epsilon^{0bmn}\partial_mB^a_{\,n}\right)&=J^{ab}+j_{(e)}^{ab}\\
		-\partial_0B^{ab}-\tfrac{1}{2}\tfrac{3\theta}{a}\left(\epsilon^{0amn}\partial_mE^b_{\,n}+\epsilon^{0bmn}\partial_mE^a_{\,n}\right)&=K^{ab}+j_{(m)}^{ab}	\ .
		\end{align}
If we take a double divergence $\partial_a\partial_b$ of these last two equations, use the electric and magnetic Gauss constraints \eqref{eGauss} and \eqref{mGauss}, and remember that the intrinsic current $j_{(e,m)}$ ave zero double divergence \eqref{j=0}, we get
		\begin{align}
		\partial_0\rho_{(e)}+\partial_a\partial_bJ^{ab}&=0\\
		\partial_0\rho_{(m)}+\partial_a\partial_bK^{ab}&=0\ .
		\end{align}
These are fractonic continuity equations for both electric and magnetic-like charges $\rho_{(e,m)}$, which implements dipole-like conservations \cite{Pretko:2016lgv,Pretko:2016kxt,Pretko:2017xar,Pretko:2020cko,Bertolini:2022ijb}. Notice that the currents $j^{ab}_{(e)}(x)$ \eqref{je} and $j^{ab}_{(m)}(x)$ \eqref{jm} does not contribute to a fractonic continuity equation, as a consequence of \eqref{j=0}.
	\ei
 If we consider a duality transformation analogous to the standard electromagnetic case \eqref{AVdual}
	\be
	A_{\mu\nu}\to-V_{\mu\nu}\quad;\quad V_{\mu\nu}\to A_{\mu\nu}\ ,
	\ee
we have
	\be
	E^{ab}\to B^{ab}\quad;\quad B^{ab}\to-E^{ab}\quad;\quad j_{(e)}\to j_{(m)}\quad;\quad j_{(m)}\to- j_{(e)}\ ,
	\ee
and the EoM are preserved provided that 
	\be\label{dual constr}
	a=-3\theta=-b\ ,
	\ee
for which
		\begin{align}
		E^{ab}&=3\theta F^{ab0}+\theta\left(\epsilon^{0amn}G^b_{\ mn}+\epsilon^{0bmn}G^a_{\ mn}\right)\label{fracE}\\
		B^{ab}&=-3\theta G^{ab0}+\theta\left(\epsilon^{0amn}F^b_{\ mn}+\epsilon^{0bmn}F^a_{\ mn}\right)\label{fracB}\ .
		\end{align}
The duality is also extended to the action, which remains invariant, by requiring $\theta \rightarrow -\theta$, which implies $\theta=\pi$, similarly to the EM case of Section \ref{EMduality}. Analogously, when matter is introduced we must also have
	\be\label{J->K-fract}
	J^{\mu\nu}\to-K^{\mu\nu}\quad;\quad K^{\mu\nu}\to J^{\mu\nu}\ .
	\ee
The action thus is
	\be\label{Sself-dual-frac}
	S_{tot}=S+S_J=\int d^4x\left[3\pi \left(-\tfrac{1}{6}F^{\mu\nu\rho}F_{\mu\nu\rho}+\tfrac{1}{6}G^{\mu\nu\rho}G_{\mu\nu\rho}+\tfrac{2}{9}\epsilon^{\mu\nu\rho\sigma}F^\lambda_{\ \mu\nu}G_{\lambda\rho\sigma}\right)-A_{\mu\nu}J^{\mu\nu}-V_{\mu\nu}K^{\mu\nu}\right] \ ,
	\ee
and the corresponding EoM are the following
		\begin{align}
		\partial_a\partial_bE^{ab}&=\rho_{(e)}\label{eGauss'}\\
		\partial_a\partial_bB^{ab}&=\rho_{(m)}\label{mGauss'}\\
		-\partial_0 E^{ab}-\tfrac{1}{2}\left(\epsilon^{0amn}\partial_mB^b_{\,n}+\epsilon^{0bmn}\partial_mB^a_{\,n}\right)&=J^{ab}+j_{(e)}^{ab}\\
		-\partial_0B^{ab}+\tfrac{1}{2}\left(\epsilon^{0amn}\partial_mE^b_{\,n}+\epsilon^{0bmn}\partial_mE^a_{\,n}\right)&=K^{ab}+j_{(m)}^{ab}	\ .
		\end{align}

\subsection{Fractonic axion-like terms : no Witten effect}\label{WittenFrac}

As discussed in the generalized Maxwell theory of Section \ref{wittenMax}, we now introduce two axion-like fractonic theta terms \cite{Bertolini:2022ijb,Bertolini:2023sqa,Bertolini:2024yur} to the action $S$ \eqref{Sfracton-dual} where the constants have already been set to the duality constraint \eqref{dual constr}. The total action is the following
	\begin{align}
	S^{(f)}_{inv}&=S+S_{\theta_1}+S_{\theta_2}\label{fract-Sem-dual-theta2}\\
	&=\int d^4x\left[3\theta \left(-\tfrac{1}{6}F^{\mu\nu\rho}F_{\mu\nu\rho}+\tfrac{1}{6}G^{\mu\nu\rho}G_{\mu\nu\rho}+\tfrac{2}{9}\epsilon^{\mu\nu\rho\sigma}F^\lambda_{\ \mu\nu}G_{\lambda\rho\sigma}\right)+\right.\nonumber\\
	&\qquad\qquad\left.+\tfrac{\theta_1}{9}\epsilon^{\mu\nu\rho\sigma}F^\lambda_{\ \mu\nu}F_{\lambda\rho\sigma}+\tfrac{\theta_2}{9}\epsilon^{\mu\nu\rho\sigma}G^\lambda_{\ \mu\nu}G_{\lambda\rho\sigma}\right] \ ,\nonumber
	\end{align}
with 
	\begin{align}
	S_{\theta_1}&\equiv\frac{1}{9}\int d^4x\,\tilde\theta_1\epsilon^{\mu\nu\rho\sigma}F^\lambda_{\ \mu\nu}F_{\lambda\rho\sigma}=\int d^4x\,\tilde\theta_1\epsilon^{\mu\nu\rho\sigma}\partial_\mu A_{\nu}^\lambda\partial_\rho A_{\lambda\sigma}\label{Sfract-theta1}\\
	S_{\theta_2}&\equiv\frac{1}{9}\int d^4x\,\tilde\theta_2\epsilon^{\mu\nu\rho\sigma}G^\lambda_{\ \mu\nu}G_{\lambda\rho\sigma}=\int d^4x\,\tilde\theta_2\epsilon^{\mu\nu\rho\sigma}\partial_\mu V_{\nu}^\lambda\partial_\rho V_{\lambda\sigma}\ ,\label{Sfract-theta2}
	\end{align}
and
	\be
	\theta=\pi\quad;\quad\theta_{1,2}=\theta_{1,2}(x)\ .
	\ee
The EoM are modified as follows
	\begin{align}
	\frac{\delta S^{(f)}_{inv}}{\delta A_{\alpha\beta}}&=-3\theta\partial_\mu F^{\alpha\beta\mu}+\frac{1}{3}\partial_\mu\theta_1\left(\epsilon^{\alpha\mu\nu\rho}F^\beta_{\ \nu\rho}+\epsilon^{\beta\mu\nu\rho}F^\alpha_{\ \nu\rho}\right)\\
		&=-3\theta\partial_\mu F^{\alpha\beta\mu}+\partial_\mu\theta_1\left(\epsilon^{\alpha\mu\nu\rho}\partial_\nu A^\beta_{\rho}+\epsilon^{\beta\mu\nu\rho}\partial_\nu A^\alpha_{\rho}\right)\\
	\frac{\delta S^{(f)}_{inv}}{\delta V_{\alpha\beta}}&=3\theta\partial_\mu G^{\alpha\beta\mu}+\frac{1}{3}\partial_\mu\theta_1\left(\epsilon^{\alpha\mu\nu\rho}G^\beta_{\ \nu\rho}+\epsilon^{\beta\mu\nu\rho}G^\alpha_{\ \nu\rho}\right)\\
		&=3\theta\partial_\mu G^{\alpha\beta\mu}+\partial_\mu\theta_1\left(\epsilon^{\alpha\mu\nu\rho}\partial_\nu V^\beta_{\rho}+\epsilon^{\beta\mu\nu\rho}\partial_\nu V^\alpha_{\rho}\right)\ .
	\end{align}
Concerning the components of these EoM we have that
	\bi
	\item {$\pmb{\alpha=\beta=0}$}
		\begin{align}
		\frac{\delta S^{(f)}_{inv}}{\delta A_{00}}&=-3\theta\partial_m F^{00m}+\frac{2}{3}\partial_m\theta_1\epsilon^{0mnp}F^0_{\ np}\\
			&=-6\theta\partial_m\left(\partial^0A^{m0}-\partial^mA^{00}\right)+2\partial_m\theta_1\epsilon^{0mnp}\partial_n A^0_{p}\\
		\frac{\delta S^{(f)}_{inv}}{\delta V_{00}}&=3\theta\partial_m G^{00m}+\frac{2}{3}\partial_m\theta_1\epsilon^{0mnp}G^0_{\ np}\\
			&=6\partial_m\left(\partial^0V^{m0}-\partial^mV^{00}\right)+2\partial_m\theta_1\epsilon^{0mnp}\partial_n V^0_{p}\ ,
		\end{align}
for which we notice that the on-shell fractonic solutions \eqref{A0} and \eqref{V0} are still valid and enforced by the theta-terms. Thus the additional properties of the field strengths \eqref{F00i=0}-\eqref{Gij0=-2G0ij} hold, and, assuming that the conjugate momenta are not changed by the introduction of the axion-like terms $S_{\theta_1}$ \eqref{Sfract-theta1} and $S_{\theta_2}$ \eqref{Sfract-theta2}, we have that the electric and magnetic -like fields are still \eqref{fracE} and \eqref{fracB}.
	\item {$\pmb{\alpha=a,\ \beta=0}$}
		\begin{align}
		\frac{\delta S^{(f)}_{inv}}{\delta A_{a0}}&=-3\theta\partial_m F^{a0m}+\partial_\mu\theta_1\cancel{\epsilon^{a\mu\nu\rho}\partial_\nu A_\rho^0}+\partial_m\theta_1\epsilon^{0mnp}\partial_nA_p^a\\
			&=\frac{1}{2}\partial_mE^{am}-\frac{\theta}{2}\epsilon^{0abc}\partial_m G^m_{\ bc}+\frac{1}{3}\partial_m\theta_1\epsilon^{0mnp}F^a_{\ np}\\
		\frac{\delta S^{(f)}_{inv}}{\delta V_{a0}}&=3\theta\partial_m G^{a0m}+\partial_\mu\theta_2\cancel{\epsilon^{a\mu\nu\rho}\partial_\nu V_\rho^0}+\partial_m\theta_2\epsilon^{0mnp}\partial_nV_p^a\\
			&=\frac{1}{2}\partial_mB^{am}-\frac{\theta}{2}\epsilon^{0abc}\partial_m F^m_{\ bc}+\frac{1}{3}\partial_m\theta_2\epsilon^{0mnp}G^a_{\ np}\ ,
		\end{align}
where we used the solutions \eqref{A0} and \eqref{V0}, and the definitions of the electric and magnetic fields \eqref{fracE} and \eqref{fracB}. Taking the divergence of the above EoM and going on-shell, we respectively get
		\begin{align}
		\partial_m\partial_nE^{mn}&=-\frac{2}{3}\partial_a\left(\epsilon^{0mnp}\partial_m\theta_1\,F^a_{\ np}\right)\label{fracMaxEom1}\\
		&=-\frac{1}{3\theta}\left[\partial_m\partial_n\theta_1\left(B^{mn}+3\theta G^{mn0}\right)+2\partial_m\theta_1\partial_n\left(B^{mn}+3\theta G^{mn0}\right)\right]\nonumber\\
		\partial_m\partial_nB^{mn}&=-\frac{2}{3}\partial_a\left(\epsilon^{0mnp}\partial_m\theta_2\,G^a_{\ np}\right)\label{fracMaxEom2}\\
		&=-\frac{1}{3\theta}\left[\partial_m\partial_n\theta_2\left(E^{mn}-3\theta F^{mn0}\right)+2\partial_m\theta_2\partial_n\left(E^{mn}-3\theta F^{mn0}\right)\right]\ ,\nonumber
		\end{align}
where we used again the definitions of electric and magnetic tensor fields \eqref{fracE} and \eqref{fracB}, and the Bianchi identity \eqref{bianchi}. We can see already that the structure is of the same kind as the one of the self-dual generalized Maxwell case of Section \ref{wittenMax}.
	\item {$\pmb{\alpha=a,\ \beta=b}$}
		\begin{align}
		\frac{\delta S^{(f)}_{inv}}{\delta A_{ab}}&=-3\theta\partial_\mu F^{ab\mu}+\frac{1}{3}\partial_\mu\theta_1\left(\epsilon^{a\mu\nu\rho}F^b_{\ \nu\rho}+\epsilon^{b\mu\nu\rho}F^a_{\ \nu\rho}\right)\\
		&=-\partial_0E^{ab}-\tfrac{1}{2}\left(\epsilon^{0amn}\partial_mB^b_n+\epsilon^{0bmn}\partial_mB^a_n\right)-j_{(e)}^{ab}+\nonumber\\
		&\quad-\frac{1}{3}\partial_0\theta_1\left(\epsilon^{0amn}F^b_{\ mn}+\epsilon^{0bmn}F^a_{\ mn}\right)-\frac{1}{2}\partial_m\theta_1\left(\epsilon^{0amn}F^b_{\ n0}+\epsilon^{0bmn}F^a_{\ n0}\right)\nonumber\\
		\frac{\delta S^{(f)}_{inv}}{\delta V_{ab}}&=3\theta\partial_\mu G^{ab\mu}+\frac{1}{3}\partial_\mu\theta_2\left(\epsilon^{a\mu\nu\rho}G^b_{\ \nu\rho}+\epsilon^{b\mu\nu\rho}G^a_{\ \nu\rho}\right)\\
		&=-\partial_0B^{ab}+\tfrac{1}{2}\left(\epsilon^{0amn}\partial_mB^b_n+\epsilon^{0bmn}\partial_mB^a_n\right)-j_{(m)}^{ab}+\nonumber\\
		&\quad-\frac{1}{3}\partial_0\theta_2\left(\epsilon^{0amn}G^b_{\ mn}+\epsilon^{0bmn}G^a_{\ mn}\right)-\frac{1}{2}\partial_m\theta_2\left(\epsilon^{0amn}G^b_{\ n0}+\epsilon^{0bmn}G^a_{\ n0}\right)\nonumber\ ,
		\end{align}
where the second line of both EoM matches exactly the Maxwell-like equations \eqref{fracMaxIntr1} and \eqref{fracMaxIntr2} respectively, while the third lines are the contributions from the axion-like theta terms. Through the definitions of the electric and magnetic -like fields, the on-shell above equations can also be rewritten, respectively, as
		\begin{align}
		0&=-\partial_0E^{ab}-\tfrac{1}{2}\left(\epsilon^{0amn}\partial_mB^b_n+\epsilon^{0bmn}\partial_mB^a_n\right)-j_{(e)}^{ab}+\label{fracMaxEom3}\\
		&\quad-\frac{1}{3\theta}\partial_0\theta_1\left(B^{ab}+3\theta G^{ab0}\right)-\frac{1}{2}\epsilon_{0kmn}\partial^m\theta_1\left(\eta^{ak}F^{bn0}+\eta^{bk}F^{an0}\right)\nonumber\\
		&=-\partial_0E^{ab}-\tfrac{1}{2}\left(\epsilon^{0amn}\partial_mB^b_n+\epsilon^{0bmn}\partial_mB^a_n\right)-j_{(e)}^{ab}+\nonumber\\
		&\quad-\frac{1}{3\theta}\partial_0\theta_1\left(B^{ab}+3\theta G^{ab0}\right)+\frac{1}{6}\partial_m\theta_1\left[\tfrac{1}{\theta}\left(\epsilon^{0amn}E_n^b+\epsilon^{0bmn}E_n^a-6\theta G^{abm}\right)-j_{(m)}^{abm}\right]\nonumber\\[10px]
		0&=-\partial_0B^{ab}+\tfrac{1}{2}\left(\epsilon^{0amn}\partial_mB^b_n+\epsilon^{0bmn}\partial_mB^a_n\right)-j_{(m)}^{ab}+\label{fracMaxEom4}\\
		&\quad-\frac{1}{3\theta}\partial_0\theta_2\left(E^{ab}-3\theta F^{ab0}\right)-\frac{1}{2}\epsilon_{0kmn}\partial^m\theta_2\left(\eta^{ak}G^{bn0}+\eta^{bk}G^{an0}\right)\nonumber\\
		&=-\partial_0B^{ab}+\tfrac{1}{2}\left(\epsilon^{0amn}\partial_mB^b_n+\epsilon^{0bmn}\partial_mB^a_n\right)-j_{(m)}^{ab}+\nonumber\\
		&\quad-\frac{1}{3\theta}\partial_0\theta_2\left(E^{ab}-3\theta F^{ab0}\right)-\frac{1}{6}\partial_m\theta_2\left[\tfrac{1}{\theta}\left(\epsilon^{0amn}B_n^b+\epsilon^{0bmn}B_n^a-6\theta F^{abm}\right)+j_{(e)}^{abm}\right]\ ,\nonumber
		\end{align}
where we defined
		\begin{align}
		j_{(m)}^{abc}&=j_{(m)}^{bac}\equiv3\left[\eta^{ab}G_n^{\ nc}-\tfrac{1}{2}\left(\eta^{bc}G_n^{\ na}+\eta^{ac}G_n^{\ nb}\right)\right]\quad;\quad\partial_cj_{(m)}^{abc}=\tfrac{2}{\theta}j_{(m)}^{ab}\\
		j_{(e)}^{abc}&=j_{(e)}^{bac}\equiv-3\left[\eta^{ab}F_n^{\ nc}-\tfrac{1}{2}\left(\eta^{bc}F_n^{\ na}+\eta^{ac}F_n^{\ nb}\right)\right]\quad;\quad\partial_cj_{(e)}^{abc}=\tfrac{2}{\theta}j_{(e)}^{ab}\ .
		\end{align}
	\ei
Similarly to the generalized self-dual Maxwell case, we now consider, for simplicity, time-dependent axion fields, $i.e.$ $\theta_{1,2}=\theta_{1,2}(t)$, for which the above fractonic equations \eqref{fracMaxEom1}, \eqref{fracMaxEom2}, \eqref{fracMaxEom3} and \eqref{fracMaxEom4} become
		\begin{align}
		\partial_m\partial_nE^{mn}&=0\\
		\partial_m\partial_nB^{mn}&=0\\
		-\partial_0E^{ab}-\tfrac{1}{2}\left(\epsilon^{0amn}\partial_mB^b_n+\epsilon^{0bmn}\partial_mB^a_n\right)&=\frac{1}{3\theta}\partial_0\theta_1\left(B^{ab}+3\theta G^{ab0}\right)+j_{(e)}^{ab}\\
		-\partial_0B^{ab}+\tfrac{1}{2}\left(\epsilon^{0amn}\partial_mB^b_n+\epsilon^{0bmn}\partial_mB^a_n\right)&=\frac{1}{3\theta}\partial_0\theta_2\left(E^{ab}-3\theta F^{ab0}\right)+j_{(m)}^{ab}\ .
		\end{align}
By taking a double derivative of the Amp\`ere and Faraday -like equations,
we can see that also in this fractonic model there is no Witten effect.

\section{Conclusions and outlook}\label{concl}

In this work, we have presented a novel tensorial generalization of EM-like duality for covariant fractons within a doubled-potential framework. The model incorporates two independent symmetric tensor gauge fields either related to the electric or to the magnetic sector. The introduction of a fractonic \textit{mutual-theta term} $S_{\theta}$ \eqref{SthetaAV} in the invariant action allows for the duality to extend also to the conjugate momenta, in the same way as for the standard dual Maxwell case studied in Section \ref{EMduality}. Thus for both electromagnetism and fractons, a properly tuned action, identified either by $S_{tot}$ \eqref{Sem-dual'+source} or $S^{(f)}_{tot}$ \eqref{Sself-dual-frac}, provides Maxwell (-like) equations, self-dual conjugate momenta, and positive-definite energy. Therefore we can claim that since the action $S^{(f)}_{tot}$ \eqref{Sself-dual-frac} with and without matter currents is invariant under the exchange of the electric-like and magnetic-like sectors, the theory exhibits a genuine fractonic self-duality. In particular the main feature of the model described by the action $S^{(f)}_{tot}$ \eqref{Sself-dual-frac} is that it provides a self-consistent and covariant description of electric-like and magnetic-like charges, introducing the concept of covariant magnetic fractons, {\it i.e.} magnetic quasiparticles with restricted mobility. This is a novelty in the context of fractons, since typically magnetic fractonic charges were introduced by hand in the EoM as vectorial quantities \cite{Pretko:2016lgv,Pretko:2016kxt}, while here they naturally arise as sources for the ``magnetic'' gauge field $V_{\mu\nu}(x)$. Moreover, differently from the EM-like duality already studied in foliated fracton phases \cite{Ma:2017aog,Shirley:2018nhn,Hsin:2021mjn,Geng:2021cmq} and in fracton magnetohydrodynamics \cite{Qi:2022seq}, our quantum field theoretical approach completely preserves the Lorentz invariance of the original single-potential covariant fractons \cite{Blasi:2022mbl,Bertolini:2022ijb,Bertolini:2023juh}. The main difference from the so called ``scalar charge theory of fractons'' \cite{Pretko:2016lgv,Pretko:2016kxt} and its covariantization \cite{Bertolini:2022ijb} is that here, as a consequence of the duality, the magnetic tensor field $B^{ij}(x)$ \eqref{fracB} turns out to be symmetric, thus this model also represents a variation of the scalar charge theory of fractons displaying magnetic scalar charges: a \textit{self-dual scalar charge theory of fractons}. Another key result of our work is the absence of a Witten-like effect \cite{Witten:1979ey,Rosenberg:2010ia}, in contrast to the non-dual case \cite{Pretko:2017xar}. As a consequence of our ``doubled'' theory, while the Witten effect still holds in the Bunster-Henneaux approach \cite{Bunster:2011qp} it is absent in our case as shown in Section \ref{wittenMax} and \ref{WittenFrac}. The magnetic charges, in our models, appear as a consequence of the additional gauge fields ($V_\mu(x)$ and $V_{\mu\nu}(x)$). We thus have a theory that when coupled to matter display two independent charges, an electric and a magnetic one. Hence more degrees of freedom than standard cases, and in terms of which the self-duality can be extended, through \eqref{J->K} and \eqref{J->K-fract}. This is one of the features that distinguishes the results of our paper from \cite{Bunster:2011qp} and similar self-dual models \cite{Schwarz:1993vs}, for which said property cannot be extended to matter sources. Therefore fractonic dyons \cite{Williamson:2018ofu,Hsin:2023ooo} do not naturally arise within the covariant doubled-potential framework.\\

Our results open new avenues for exploring the interplay between fracton physics, dualities, and relativistic field theories. The doubled-potential framework enriches our understanding of gauge theories with higher-rank tensorial fields and provides a stepping stone toward a deeper understanding of fractonic excitations in covariant settings. Looking ahead, several promising directions emerge. Firstly, extending this framework to interacting systems could uncover new aspects of fractonic dynamics and their implications for quantum many-body systems with constrained mobility. Secondly, the connection between our covariant fracton model and extensions of linearized gravity hints at potential applications in exploring novel gravitational phenomena. In particular, the implications of fractonic self-duality for bi-metric gravitational theories \cite{Hassan:2011zd,Bunster:2013tc} warrant further study. Moreover, the study of possible topological phases and defect structures in systems governed by this doubled-potential framework could lead to new insights into the interplay between topology and restricted mobility in quantum matter. In particular, the mutual theta term in our theory naturally gives rise to a fractonic BF-like term \cite{Bertolini:2025jul} on the boundary, which can encode some topological features of the boundary states. Finally, the possible relationship between our covariant fracton framework in four dimensions and other known dualities in lower-dimensional field theories poses intriguing questions about a unified perspective on dualities across dimensions and contexts. In fact,  in condensed matter, the EM duality is linked to the particle-vortex duality of three-dimensional topological insulators \cite{Metlitski:2015eka}. Thus, the fractonic self-duality derived in our work could shed some light on the possible existence of a fractonic particle-vortex duality in topological matter.  By bridging the gap between fracton physics, gauge theories, and gravitational extensions, this work lays the foundation for theoretical developments and potential applications in both high-energy physics and condensed matter systems.

\section*{Acknowledgments}
The authors thank Nicola Maggiore for enlightening discussions.

\appendix

\section{Linearized bi-metric theory and fractons}\label{appA}
Linearized gravity is a well-established approximation of Einstein's general relativity, where the spacetime metric is treated as a small perturbation around a fixed background, typically flat Minkowski space. This framework is invaluable for studying weak-field gravitational phenomena, such as gravitational waves, and serves as a cornerstone for exploring connections between gravity and quantum field theory. Here, we propose a generalization of linearized gravity by introducing two independent spin-2 fields, effectively doubling the degrees of freedom of gravity. This approach mirrors the doubled-potential framework for gauge fields discussed previously and allows for a symmetric treatment of "electric-like" and "magnetic-like" sectors of the gravitational field. The two independent tensor fields can be interpreted as representing distinct sectors that may encode additional symmetries, constraints, or novel gravitational interactions. However, for simplicity, we will consider here the two gravitational sectors only coupled via the mutual theta-like term already previously introduced for the covariant magnetic fractons. The corresponding action that takes into account both the double-potential theory of covariant fractons together with a linearized bi-metric theory is given by
\be
S_{2}=
\int d^4x\left(\tfrac{a}{6}F^{\mu\nu\rho}F_{\mu\nu\rho}+\tfrac{b}{6}G^{\mu\nu\rho}G_{\mu\nu\rho}+\tfrac{a_1}{4}F^\mu_{\ \mu\nu} F_\rho^{\ \rho\nu}+\tfrac{b_1}{4}G^\mu_{\ \mu\nu} G_\rho^{\ \rho\nu}+\tfrac{2}{3}\theta\epsilon^{\mu\nu\rho\sigma}F^\lambda_{\ \mu\nu}G_{\lambda\rho\sigma}\right)
\label{S2}\ ,
\ee
which reduces to LG when $a=-a_1$ and $b=-b_1$ and thus to a theory that is invariant under standard infinitesimal diffeomorphisms, and to the traceless case, for which the theory only depends on traceless parts $\bar A_{\mu\nu}(x)\ ,\ \bar V_{\mu\nu}(x)$ of $A_{\mu\nu}(x)$ and $V_{\mu\nu}(x)$  when $a=-3a_1$ and $b=-3b_1$, in which case the theory is invariant under the traceless covariant fracton symmetry \cite{Prem:2017kxc,Du:2021pbc,Blasi:2022mbl,Bertolini:2022ijb,Bertolini:2023juh,Bertolini:2024yur}
	\be
	\bar\delta\bar A_{\mu\nu}=\partial_\mu\partial_\nu\lambda-\frac{1}{3}\eta_{\mu\nu}\partial^2\lambda\quad;\quad\bar\delta' \bar V_{\mu\nu}=\partial_\mu\partial_\nu\lambda'-\frac{1}{3}\eta_{\mu\nu}\partial^2\lambda'\ .
	\ee
Equations of motion
	\begin{align}
	\frac{\delta S_2}{\delta A_{\alpha\beta}}&=a\partial_\mu F^{\alpha\beta\mu}+\tfrac{a_1}{2}\left(2\eta^{\alpha\beta}\partial_\lambda F_\mu^{\ \mu\lambda}-\partial^\alpha F_\mu^{\ \mu\beta}-\partial^\beta F_\mu^{\ \mu\alpha}\right)\label{EoMA2}\\
	\frac{\delta S_2}{\delta V_{\alpha\beta}}&=b\partial_\mu G^{\alpha\beta\mu}+\tfrac{b_1}{2}\left(2\eta^{\alpha\beta}\partial_\lambda G_\mu^{\ \mu\lambda}-\partial^\alpha G_\mu^{\ \mu\beta}-\partial^\beta G_\mu^{\ \mu\alpha}\right)\ .\label{EoMV2}
	\end{align}
Conjugate momenta
	\begin{align}
	\frac{\delta S_2}{\delta\partial_0A_{\alpha\beta}}&=\Pi_{(\textsc{a})}^{\alpha\beta}\\
	&=-aF^{\alpha\beta0}-\tfrac{a_1}{2}\left(2\eta^{\alpha\beta}F_m^{\ m0}-\eta^{\alpha0}F_\lambda^{\ \lambda\beta}-\eta^{\beta0}F_\lambda^{\ \lambda\alpha}\right)+\theta\left(\epsilon^{0\alpha mn}G^\beta_{\ mn}+\epsilon^{0\beta mn}G^\alpha_{\ mn}\right)\nonumber\\
	\frac{\delta S_2}{\delta\partial_0V_{\alpha\beta}}&=\Pi_{(\textsc{v})}^{\alpha\beta}\\
	&=-bG^{\alpha\beta0}-\tfrac{b_1}{2}\left(2\eta^{\alpha\beta}G_m^{\ m0}-\eta^{\alpha0}G_\lambda^{\ \lambda\beta}-\eta^{\beta0}G_\lambda^{\ \lambda\alpha}\right)+\theta\left(\epsilon^{0\alpha mn}F^\beta_{\ mn}+\epsilon^{0\beta mn}F^\alpha_{\ mn}\right)\ ,\nonumber
	\end{align}
thus
	\begin{align}
	\Pi_{(\textsc{a})}^{00}&=\Pi_{(\textsc{v})}^{00}=0\\
	\Pi_{(\textsc{a})}^{a0}&=\tfrac{1}{2}(a+a_1)F^{00a}-\tfrac{a_1}{2}F_m^{\ ma}-\theta\epsilon^{0amn}G_{0mn}\\
	\Pi_{(\textsc{v})}^{a0}&=\tfrac{1}{2}(b+b_1)G^{00a}-\tfrac{b_1}{2}G_m^{\ ma}-\theta\epsilon^{0amn}F_{0mn}\\
	\Pi_{(\textsc{a})}^{ab}&=-aF^{ab0}-a_1\eta^{ab}F_m^{\ m0}+\theta\left(\epsilon^{0amn}G^b_{\ mn}+\epsilon^{0bmn}G^a_{\ mn}\right)\\
	\Pi_{(\textsc{v})}^{ab}&=-bG^{ab0}-b_1\eta^{ab}G_m^{\ m0}+\theta\left(\epsilon^{0amn}F^b_{\ mn}+\epsilon^{0bmn}F^a_{\ mn}\right)\ .
	\end{align}
Components of the EoM:
	\bi
	\item $\pmb{\alpha=\beta=0\ :}$
		\begin{align}
		\partial_m\left[(a+a_1) F^{00m}-a_1F_n^{\ nm}\right]&=2\partial_m\Pi_{(\textsc{a})}^{m0}\label{EoMA200}\\
		\partial_m\left[(b+b_1) G^{00m}-b_1G_n^{\ nm}\right]&=2\partial_m\Pi_{(\textsc{v})}^{m0}\ .\label{EoMV200}
		\end{align}
On shell these imply
		\begin{align}
		2\partial_m\Pi_{(\textsc{a})}^{m0}&=0\label{EoMA200os}\\
		2\partial_m\Pi_{(\textsc{v})}^{m0}&=0\ .\label{EoMV200os}
		\end{align}
	\item $\pmb{\alpha=a\ ,\ \beta=0\ :}$ on-shell gives
		\begin{align}
		0&=a\partial_\mu F^{a0\mu}-\tfrac{a_1}{2}\left(\partial^aF_m^{\ m0}+\partial^0F_\mu^{\ \mu a}\right)\label{EoMA2a0}\\
		&=-\tfrac{1}{2}(a+a_1)\partial_0F^{00a}+\tfrac{a_1}{2}\partial_0F_m^{\ ma}+a\partial_mF^{0am}-\tfrac{a_1}{2}\partial^aF_m^{\ m0}\nonumber\\
		&=-\partial_0\Pi_{(\textsc{a})}^{a0}-\theta\epsilon^{0amn}\partial_0G_{0mn}+a\partial_mF^{0am}-\tfrac{a_1}{2}\partial^aF_m^{\ m0}\nonumber\\
		&=-\partial_0\Pi_{(\textsc{a})}^{a0}-\tfrac{3}{2}\theta\epsilon^{0amn}\partial_mG_{00n}+a\partial_mF^{0am}-\tfrac{a_1}{2}\partial^aF_m^{\ m0}\nonumber\\
		0&=b\partial_\mu G^{a0\mu}-\tfrac{b_1}{2}\left(\partial^aG_m^{\ m0}+\partial^0G_\mu^{\ \mu a}\right)\label{EoMA2a0}\\
		&=-\tfrac{1}{2}(b+b_1)\partial_0G^{00a}+\tfrac{b_1}{2}\partial_0G_m^{\ ma}+b\partial_mG^{0am}-\tfrac{b_1}{2}\partial^aG_m^{\ m0}\nonumber\\
		&=-\partial_0\Pi_{(\textsc{v})}^{a0}-\theta\epsilon^{0amn}\partial_0F_{0mn}+b\partial_mG^{0am}-\tfrac{b_1}{2}\partial^aG_m^{\ m0}\nonumber\\
		&=-\partial_0\Pi_{(\textsc{v})}^{a0}-\tfrac{3}{2}\theta\epsilon^{0amn}\partial_mF_{00n}+b\partial_mG^{0am}-\tfrac{b_1}{2}\partial^aG_m^{\ m0}\ .\nonumber
		\end{align}
where we used the implication of the Bianchi identity
		\be
		\theta\epsilon^{0amn}\partial_0G_{0mn}=\tfrac{3}{2}\theta\epsilon^{0amn}\partial_mG_{00n}
		\ee
and same for $F$. Taking the divergence $\partial_a$ of these EoM we get
		\begin{align}
		0&=-\partial_0\cancel{\partial_a\Pi_{(\textsc{a})}^{a0}}-\tfrac{3}{2}\theta\bcancel{\epsilon^{0amn}\partial_a\partial_m}G_{00n}+a\partial_a\partial_bF^{0ab}-\tfrac{a_1}{2}\partial^2F_m^{\ m0}\nonumber\\
		&=-\tfrac{a}{2}\partial_a\partial_bF^{ab0}-\tfrac{a_1}{2}\partial^2F_m^{\ m0}\nonumber\\
		&=\tfrac{1}{2}\partial_a\partial_b\Pi_{(\textsc{a})}^{ab}\\
		0&=-\partial_0\cancel{\partial_a\Pi_{(\textsc{v})}^{a0}}-\tfrac{3}{2}\theta\bcancel{\epsilon^{0amn}\partial_a\partial_m}F_{00n}+b\partial_a\partial_vG^{0ab}-\tfrac{b_1}{2}\partial^2G_m^{\ m0}\nonumber\\
		&=-\tfrac{b}{2}\partial_a\partial_bG^{ab0}-\tfrac{b_1}{2}\partial^2G_m^{\ m0}\nonumber\\
		&=\tfrac{1}{2}\partial_a\partial_b\Pi_{(\textsc{v})}^{ab}\ ,
		\end{align}
where we used the on-shell EoM \eqref{EoMA200os} and \eqref{EoMV200os} and the cyclicity of the field strengths.
	\item $\pmb{\alpha=a\ ,\ \beta=b\ :}$ from the on-shell EoM \eqref{EoMA2} we have
		\be
			\begin{split}
			0=&a\partial_\mu F^{ab\mu}+\tfrac{a_1}{2}\left(2\eta^{ab}\partial_\lambda F_\mu^{\ \mu\lambda}-\partial^aF_\mu^{\ \mu b}-\partial^b F_\mu^{\ \mu a} \right)\\
			=&\partial_0\left(aF^{ab0}+a_1\eta^{ab}F_m^{\ m0}\right)+a\partial_mF^{abm}+\tfrac{a_1}{2}\left(2\eta^{ab}\partial_n F_\mu^{\ \mu n}-\partial^aF_\mu^{\ \mu b}-\partial^b F_\mu^{\ \mu a} \right)\\
			=&-\partial_0\Pi^{ab}_{(\textsc a)}+\theta \left(\epsilon^{0amn}\partial_0G^b_{\ mn}+\epsilon^{0bmn}\partial_0G^a_{\ mn}\right)+a\partial_mF^{abm}\\
			&+\tfrac{a_1}{2}\left(2\eta^{ab}\partial_n F_\mu^{\ \mu n}-\partial^aF_\mu^{\ \mu b}-\partial^b F_\mu^{\ \mu a} \right)\\
			=&-\partial_0\Pi^{ab}_{(\textsc a)}+\theta \left[\epsilon^{0amn}\partial_m\left(G^b_{\ 0n}-G^b_{\ n0}\right)+\epsilon^{0bmn}\partial_m\left(G^a_{\ 0n}-G^a_{\ n0}\right)\right]+a\partial_mF^{abm}\\
			&+\tfrac{a_1}{2}\left[\left(2\eta^{ab}\partial_n F_m^{\ m n}-\partial^aF_m^{\ m b}-\partial^b F_m^{\ m a} \right)-\left(2\eta^{ab}\partial_n F^{00 n}-\partial^aF^{00 b}-\partial^b F^{00 a} \right)\right]\\
			=&-\partial_0\Pi^{ab}_{(\textsc a)}-\tfrac{\theta}{b}\left(\epsilon^{0amn}\partial_m\Pi^b_{(\textsc v)\,n}+\epsilon^{0bmn}\partial_m\Pi^a_{(\textsc v)\,n}\right)+\left(\tfrac{6\theta^2}{b}+a\right)\partial_mF^{abm}\\
			&+\tfrac{1}{2}\left(\tfrac{3\theta^2}{b}+a_1\right)\left(2\eta^{ab}\partial_n F_m^{\ m n}-\partial^aF_m^{\ m b}-\partial^b F_m^{\ m a} \right)-\\
			&-\tfrac{a_1}{2}\left(2\eta^{ab}\partial_n F^{00 n}-\partial^aF^{00 b}-\partial^b F^{00 a} \right)	+\theta \left(\epsilon^{0amn}\partial_mG^b_{\ 0n}+\epsilon^{0bmn}\partial_mG^a_{\ 0n}\right)
			\end{split}
		\ee
where we have used 
		\be
		\epsilon^{0amn}\partial_0G^b_{mn}=\epsilon^{0amn}\partial_m\left(G^a_{\ 0n}-G^a_{n0}\right)
		\ee
from the Bianchi identity, and
		\be\label{epsilonPi}
			\begin{split}
			\epsilon^{0amn}\Pi^b_{(\textsc v)\,n}+\epsilon^{0bmn}\Pi^a_{(\textsc v)\,n}=&b\left(\epsilon^{0amn}\partial_m G^a_{\ n0}+\epsilon^{0bmn}\partial_m G^a_{\ n0}\right)+6\theta\partial_mF^{abm}+\\
			&+\tfrac{3}{2}\theta\left(2\eta^{ab}\partial_mF_n^{\ nm}-\partial^aF_n^{\ nb}-\partial^bF_n^{\ na}\right)
			\end{split}
		\ee
due to cyclicity \eqref{cicl} and properties of Levi-Civita products \eqref{eps1} and \eqref{eps2}. In the same way, we also have, from \eqref{EoMV2}
		\be
			\begin{split}
			0=&-\partial_0\Pi^{ab}_{(\textsc v)}-\tfrac{\theta}{a}\left(\epsilon^{0amn}\partial_m\Pi^b_{(\textsc a)\,n}+\epsilon^{0bmn}\partial_m\Pi^a_{(\textsc a)\,n}\right)+\left(\tfrac{6\theta^2}{a}+b\right)\partial_mF^{abm}\\
			&+\tfrac{1}{2}\left(\tfrac{3\theta^2}{a}+b_1\right)\left(2\eta^{ab}\partial_n F_m^{\ m n}-\partial^aF_m^{\ m b}-\partial^b F_m^{\ m a} \right)-\\
			&-\tfrac{b_1}{2}\left(2\eta^{ab}\partial_n F^{00 n}-\partial^aF^{00 b}-\partial^b F^{00 a} \right)+\theta \left(\epsilon^{0amn}\partial_mG^b_{\ 0n}+\epsilon^{0bmn}\partial_mG^a_{\ 0n}\right)
			\end{split}
		\ee
	\ei
In order to establish a link with fractonic Maxwell-like features, it is necessary to require \eqref{A0} and \eqref{V0}, $i.e.$
	\begin{equation}\label{A02}
	A_{0\mu}=A_{\mu0}\equiv\partial_\mu A_0\quad;\quad V_{0\mu}=V_{\mu0}\equiv\partial_\mu V_0
	\end{equation}
such that
	\begin{empheq}{align}
	&F^{i00}=F^{0i0}=F^{00i}=0\label{F00i=02}\\
	&G^{i00}=G^{0i0}=G^{00i}=0\label{G00i=02}\\
	&F^{ij0}=-2F^{0ij}=-2F^{i0j}\label{Fij0=-2F0ij2}\\
	&G^{ij0}=-2G^{0ij}=-2G^{i0j}\label{Gij0=-2G0ij2}\ ,
	\end{empheq}
which however are not solutions to the EoM anymore, and should thus be imposed fo instance through the introduction of a Lagrange multiplier. Under these conditions all the above momenta and on-shell EoM becomes
	\begin{align}
	\Pi_{(\textsc{a})}^{00}&=\Pi_{(\textsc{v})}^{00}=0\\
	\Pi_{(\textsc{a})}^{a0}&=-\tfrac{a_1}{2}F_m^{\ ma}\\
	\Pi_{(\textsc{v})}^{a0}&=-\tfrac{b_1}{2}G_m^{\ ma}\\
	\Pi_{(\textsc{a})}^{ab}&=-aF^{ab0}-a_1\eta^{ab}F_m^{\ m0}+\theta\left(\epsilon^{0amn}G^b_{\ mn}+\epsilon^{0bmn}G^a_{\ mn}\right)\\
	\Pi_{(\textsc{v})}^{ab}&=-bG^{ab0}-b_1\eta^{ab}G_m^{\ m0}+\theta\left(\epsilon^{0amn}F^b_{\ mn}+\epsilon^{0bmn}F^a_{\ mn}\right)\ .
	\end{align}
	\bi
	\item $\pmb{\alpha=\beta=0\ :}$
		\begin{align}
		-a_1\partial_mF_n^{\ nm}&=2\partial_m\Pi_{(\textsc{a})}^{m0}=0\label{EoMA200sol}\\
		-b_1\partial_mG_n^{\ nm}&=2\partial_m\Pi_{(\textsc{v})}^{m0}=0\ .\label{EoMV200sol}
		\end{align}
\item $\pmb{\alpha=a\ ,\ \beta=0\ :}$
		\begin{align}
		0&=-\partial_0\Pi_{(\textsc{a})}^{a0}+a\partial_mF^{0am}-\tfrac{a_1}{2}\partial^aF_m^{\ m0}\nonumber\\
		0&=-\partial_0\Pi_{(\textsc{v})}^{a0}+b\partial_mG^{0am}-\tfrac{b_1}{2}\partial^aG_m^{\ m0}\ ,\nonumber
		\end{align}
for which, taking the divergence $\partial_a$, gives, again
		\begin{align}
		\tfrac{1}{2}\partial_a\partial_b\Pi_{(\textsc{a})}^{ab}&=0\\
		\tfrac{1}{2}\partial_a\partial_b\Pi_{(\textsc{v})}^{ab}&=0\ .
		\end{align}
	\item $\pmb{\alpha=a\ ,\ \beta=b\ :}$
		\be
			\begin{split}
			0=&-\partial_0\Pi^{ab}_{(\textsc a)}-\tfrac{3\theta}{2b}\left(\epsilon^{0amn}\partial_m\Pi^b_{(\textsc v)\,n}+\epsilon^{0bmn}\partial_m\Pi^a_{(\textsc v)\,n}\right)+\left(\tfrac{9\theta^2}{b}+a\right)\partial_mF^{abm}\\
			&+\tfrac{1}{2}\left(\tfrac{9\theta^2}{2b}+a_1\right)\left(2\eta^{ab}\partial_n F_m^{\ m n}-\partial^aF_m^{\ m b}-\partial^b F_m^{\ m a} \right)\\
			0=&-\partial_0\Pi^{ab}_{(\textsc v)}-\tfrac{3\theta}{2a}\left(\epsilon^{0amn}\partial_m\Pi^b_{(\textsc a)\,n}+\epsilon^{0bmn}\partial_m\Pi^a_{(\textsc a)\,n}\right)+\left(\tfrac{9\theta^2}{a}+b\right)\partial_mG^{abm}\\
			&+\tfrac{1}{2}\left(\tfrac{9\theta^2}{2a}+b_1\right)\left(2\eta^{ab}\partial_n G_m^{\ m n}-\partial^aG_m^{\ m b}-\partial^b G_m^{\ m a} \right)\ ,
			\end{split}
		\ee
where the difference in some numerical factors are due to the fact that now there is an additional contribution of \eqref{epsilonPi} due to $G_{0mn}=-\tfrac{1}{2}G_{mn0}$. If we thus tune
		\be\label{par2}
		b=-\frac{9\theta^2}{a}=2b_1\quad;\quad a_1=\frac{1}{2}a
		\ee 
we get
		\be
			\begin{split}
			-\partial_0\Pi^{ab}_{(\textsc a)}+\tfrac{1}{2}\tfrac{a}{3\theta}\left(\epsilon^{0amn}\partial_m\Pi^b_{(\textsc v)\,n}+\epsilon^{0bmn}\partial_m\Pi^a_{(\textsc v)\,n}\right)&=0\\
			-\partial_0\Pi^{ab}_{(\textsc v)}-\tfrac{1}{2}\tfrac{3\theta}{a}\left(\epsilon^{0amn}\partial_m\Pi^b_{(\textsc a)\,n}+\epsilon^{0bmn}\partial_m\Pi^a_{(\textsc a)\,n}\right)&=0\ ,
			\end{split}
		\ee
and we can finally set
		\be\label{par2}
		a=-3\theta\quad\Rightarrow\quad b=3\theta=2b_1\quad;\quad a_1=-\frac{3}{2}\theta
		\ee 
and define
		\be
		E^{ab}\equiv\Pi^{ab}_{(\textsc a)}\quad;\quad B^{ab}\equiv\Pi^{ab}_{(\textsc v)}
		\ee
in order to have fractonic equations
		\be
			\begin{split}
			-\partial_0E^{ab}-\left(\epsilon^{0amn}\partial_mB^b_{\,n}+\epsilon^{0bmn}\partial_mB^a_{\,n}\right)&=0\\
			-\partial_0B^{ab}+\left(\epsilon^{0amn}\partial_mE^b_{\,n}+\epsilon^{0bmn}\partial_mE^a_{\,n}\right)&=0\ ,
			\end{split}
		\ee
	\ei
With the choice \eqref{par2} of the parameters, the action is the following
\be
S_{2}=3\theta
\int d^4x\left(-\tfrac{1}{6}F^{\mu\nu\rho}F_{\mu\nu\rho}+\tfrac{1}{6}G^{\mu\nu\rho}G_{\mu\nu\rho}-\tfrac{1}{8} F^\mu_{\ \mu\nu} F_\rho^{\ \rho\nu}+\tfrac{1}{8}G^\mu_{\ \mu\nu} G_\rho^{\ \rho\nu}+\tfrac{2}{9}\epsilon^{\mu\nu\rho\sigma}F^\lambda_{\ \mu\nu}G_{\lambda\rho\sigma}\right).
\label{S2fixsol}
\ee

\printbibliography[heading=bibintoc]

\end{document}